\documentclass{article}

\usepackage{arXiv}

\usepackage[utf8]{inputenc} 
\usepackage[T1]{fontenc}    
\usepackage{hyperref}       
\usepackage{url}            
\usepackage{booktabs}       
\usepackage{amsfonts}       
\usepackage{nicefrac}       
\usepackage{microtype}      
\usepackage{lipsum}		
\usepackage{graphicx}
\usepackage{doi}

\usepackage[numbers,sort&compress]{natbib}
\title{Waves in ice}

\author{Luke G.~Bennetts \\
	University of Melbourne, Australia \\
	\texttt{luke.bennetts@unimelb.edu.au} 
}

\date{}

\renewcommand{\shorttitle}{Waves in ice}

\hypersetup{
pdftitle={Waves in ice},
pdfsubject={},
pdfauthor={Luke G.~Bennetts},
pdfkeywords={},
}

\usepackage{amsmath}

\newcommand{\imag}{\textrm{Im}}

\renewcommand{\d}{\,\mathrm{d}} 	

\newcommand{\wrt}{\,\mathrm{d}}   
    
\newcommand{\ci}{\mathrm{i}}   
\newcommand{\e}{\mathrm{e}}

\usepackage{enumerate}

\usepackage{overpic,rotating}

\newcommand{\beq}{\begin{equation}}
\newcommand{\eeq}{\end{equation}}
\newcommand{\bea}{\begin{eqnarray}}
\newcommand{\eea}{\end{eqnarray}}
\newcommand{\bean}{\begin{eqnarray*}}
\newcommand{\eean}{\end{eqnarray*}}

\newcommand{\deriv}[2]{\frac{{\rm d} {#1}}{{\rm d} {#2}}}

\def\D{}
\renewcommand{\D}[2]{\mathchoice
  {\frac{\partial #2}{\partial #1}}
  {{\partial #2}/{\partial #1}}
  {{\partial #2}/{\partial #1}}
  {{\partial #2}/{\partial #1}}
  }
\newcommand{\Dn}[3]{\mathchoice
  {\frac{\partial^{#2} #3}{\partial #1^{#2}}}%
  {{\partial^{#2} #3}/{\partial #1^{#2}}}%
  {{\partial^{#2} #3}/{\partial #1^{#2}}}%
  {{\partial^{#2} #3}/{\partial #1^{#2}}}%
  }
\newcommand{\DD}[2]{\Dn{#1}2{#2}}

\newlength{\mylinelength}
\newlength{\mydashlength}
\newlength{\mydashspace}
\newlength{\mychainlengthA}
\newlength{\mychainlengthB}
\newlength{\mychainspace}
\newlength{\mylinethickness}
\newcommand{\mathand}{\quad\textnormal{and}\quad}
\newcommand{\mathandR}{\textnormal{and}\quad}

\newcommand{\mathwhere}{\quad\textnormal{where}\quad}

\newcommand{\mathXL}[1]{\quad\textnormal{#1}}

\usepackage{pifont}
\ifdefined\showmylabels
 \usepackage[inline]{showlabels}
 
\fi

\usepackage[normalem]{ulem}
\usepackage{soul}
\usepackage{tcolorbox}
\ifdefined\shownewold
   \newcommand{\del}[1]{{\color{red}\sout{#1}}}
\else
   \newcommand{\del}[1]{\ignorespaces}
\fi
\ifdefined\shownewold
   \newcommand{\deleqn}[1]{\\\del{\parbox{\textwidth}{#1}}}
\else
   \newcommand{\deleqn}[1]{\ignorespaces}
\fi
\ifdefined\shownewold
   \newcommand{\new}[1]{{\color[rgb]{0,0,1}#1}}
\else
   \newcommand{\new}[1]{#1}
\fi
\ifdefined\shownewold
   \newcommand{\old}[1]{{\color[rgb]{0.5,0.5,0.5}#1}}
\else
   \newcommand{\old}[1]{\ignorespaces}
\fi
\ifdefined\shownewold
   \newcommand{\delref}[1]{\del{\parbox{\figwidth\columnwidth}{#1}}}
\else
   \newcommand{\delref}[1]{\vspace{-12pt}}
\fi
\ifdefined\shownewold
   \newcommand{\lu}[1]{{\color[rgb]{1,0,1} #1}}
   \newcommand{\luchange}[2]{\del{#1}\new{#2}}
   \newcommand{\luleft}[1]{{\color[rgb]{1,0,1}$\longleftarrow$#1}}
   \newcommand{\luright}[1]{{\color[rgb]{1,0,1}#1$\longrightarrow$}}
\else
   \newcommand{\lu}[1]{\ignorespaces}
   \newcommand{\luchange}[1]{\ignorespaces}
   \newcommand{\luleft}[1]{\ignorespaces}
   \newcommand{\luright}[1]{\ignorespaces}
\fi
\usepackage{tikz,pgfplots}
\usetikzlibrary{calc,patterns,decorations.pathmorphing,decorations.markings,decorations.pathmorphing,angles,quotes,arrows,arrows.meta}
\tikzset{snake it/.style={decorate, decoration=snake}}

\newlength{\figurewidth}
\newlength{\figwidth}
\newlength{\figureheight}

\usepackage{xcolor}
\definecolor{darkgreen}{rgb}{0,0.55,0}
\definecolor{midgreen}{rgb}{0,0.8,0.2}
\definecolor{magenta}{rgb}{1,0,1}
\definecolor{purple}{rgb}{0.5,0,0.5}
\definecolor{darkorange}{rgb}{1,0.55,0}
\definecolor{maroon}{rgb}{0.5,0,0}
\definecolor{olive}{rgb}{0.5,0.5,0}
\definecolor{midgrey}{rgb}{0.5,0.5,0.5}
\definecolor{lightgrey}{rgb}{0.75,0.75,0.75}
\definecolor{matlabblue}{rgb}{0,0.447,0.741}
\definecolor{matlabred}{rgb}{0.85,0.325,0.098}
\definecolor{lightblue}{rgb}{0,0.5,1}
\definecolor{darkgrey}{rgb}{0.25,0.25,0.25}
\definecolor{teal}{rgb}{0,0.5,0.5}
\definecolor{navy}{rgb}{0,0,0.5}
\definecolor{goldenrod}{rgb}{0.85,0.6,0.1}

\begin{document}
\maketitle

\begin{abstract}
Ocean surface waves can propagate long distances through regions containing floating ice covers. The impacts ocean waves have on the ice covers are of interest in the climate change era, as the polar regions experience pressure from rising temperatures. This chapter provides a review of observations and theoretical models for ocean wave propagation through the marginal ice zone, landfast ice and ice shelves. It traces the historical evolution of the field, from seminal work in the 1970s–80s up to recent research advances. Key research questions are identified for each of the three ice covers, and commonalities between them are highlighted. The chapter concludes with perspectives and outlooks on the field of waves in ice, in the context of the dramatic changes currently occurring to the world's sea ice and ice shelves.
\end{abstract}

\keywords{ocean waves \and wave--ice interactions \and landfast ice \and marginal ice zone \and ice shelves \and field observations \and theoretical models \and wave attenuation \and wave scattering \and wave energy dissipation \and viscoelasticity \and ice shelf vibrations}

\begin{tcolorbox}[width=\linewidth, colback=white!95!yellow] 
\begin{itemize}
    \item A synthesis of research on ocean wave propagation through landfast ice, ice shelves and the marginal ice zone 
    \item Observations and theories reviewed for waves in each of the three types of ice-covered waters
    \item Key research questions identified and overlaps highlighted between the sub-fields
    \item Future research focus on the coupled marginal ice zone--landfast ice--ice shelf system is called for
\end{itemize}
\end{tcolorbox}

%
\section{Introduction} \label{sec:intro}

The phrase ``waves in ice" has been broadly adopted 
as an abbreviation for ``ocean wave propagation through ice-covered waters".
It has become synonymous with studies of the marginal ice zone, which is the tens to hundreds of kilometres wide outer region of the sea ice-covered ocean, where surface waves from the open ocean regularly influence the ice cover by breaking up larger ice floes, preventing the ice cover from consolidating, and more \citep{bennetts2024closing}.   
Marginal ice zone dynamics has resurfaced as an area of major international research activity  \citep{bennetts2022marginal}. 
The renewed activity was initially driven by retreat of Arctic sea ice and the associated shift of the Arctic sea ice cover towards more marginal ice zone-type conditions \citep{squire2011past} but is now equally motivated by understanding the response of Antarctic sea ice to climate change. 
Substantial advances have been made, including dedicated experiments during field campaigns, including the second Sea Ice Physics and Ecosystems eXperiment (SIPEX~II) in the eastern Southern Ocean in 2012 \citep{kohout2014storm}, the SeaState campaign in the western Arctic Sea in 2015 \citep{thomson2018overview},      and the Polynyas,	Ice	Production	and	seasonal	Evolution	in	the	Ross	Sea (PIPERS) in 2017 \citep{kohout2020observations}. 
The research focus has been on understanding and modelling wave attenuation over distance due to the ice cover, with a sub-focus on how the ice cover affects the directional wave spectrum, as these inform predictions of the distribution of wave energy in the marginal ice zone, which is the basis for modelling wave impacts on the ice cover \citep{bennetts2022theory}.
The advances build on seminal work conducted in the 1970s--1980s, led by members of the University of Cambridge's Scott Polar Research Institute, which included large experimental programmes, such as the Arctic marginal ice zone experiments (MIZEX~1983, etc.) and the Labrador ice margin experiment (LIMEX), and theoretical modelling in the 1990s--2000s, primarily by New Zealand-based researchers \citep{squire2022marginal,bennetts2022marginal}.
However, key research questions remain about waves in the marginal ice zone, as well as other aspects of marginal ice zone dynamics \citep{squire2022prognosticative}.

Waves in ice also encompasses ocean wave propagation through landfast ice, which is sea ice attached to the coast, and through ice shelves (and ice tongues), which are the extensions of grounded (freshwater) ice sheets onto the ocean surface that enclose sub-shelf water cavities.
These two forms of floating ice occupy large proportions of the Antarctic coastline, and are also found around land masses in the Arctic Ocean.
The key research questions for waves in these ice types overlap with those for waves in the marginal ice zone. 
They were topics of research activity in a similar era to the early research drive on waves in the marginal ice zone, and often led by the same research groups.
They have become active research areas again over the past one to two decades, although not at the same level of research intensity as waves in the marginal ice zone.

For landfast ice, the aim is to understand and predict breakup of the ice cover due to ocean waves that reach coastal regions during lows or absence of surrounding pack ice \citep{crocker1989breakup}.
As such, there is a focus on wave attenuation over distance travelled through landfast ice.
The topic has been revisited by researchers primarily interested in the marginal ice zone, who viewed it as providing a simplified version of the attenuation problem, because ocean waves propagate through landfast ice-covered waters as ice-coupled waves known as flexural-gravity waves, whereas their form is undetermined in the granular ice covers that occupy the marginal ice zone.
There has also been a strong focus on refection of incident wave energy by the landfast ice edge and the resulting proportion of energy transmitted as flexural-gravity waves.

Ice shelves are tens to hundreds of metres thick at the shelf front, compared to decimetres to metres for sea ice, which means the ice shelf front (its seaward edge) reflects short waves, and only long ocean waves penetrate into the ice shelf.
There is strong evidence that long waves over a broad spectrum, from long swell, to infragravity waves, to tsunamis, force ice shelf flexure that triggered major calving events \citep{brunt2011antarctic,bromirski2010transoceanic,massom2018antarctic,zhao2024long}.
As such, a key research question is on how waves interact with ice shelf fractures and other weaknesses.
Large ice shelf thickness also means that multiple propagating wave modes are likely to exist, and assessing the regimes in which these modes are significant is a topic of current research interest. 

There are existing review articles on or including waves in sea ice.
The highly cited ``of ocean waves and sea ice" trilogy \citep{squire1995ocean,squire2007ocean,squire2020ocean} and one independent review article \citep{shen2019modelling} cover waves in landfast ice and waves in the marginal ice zone.
A theme issue of \textit{Philosophical Transactions A} on marginal ice zone dynamics \citep{bennetts2022marginal} includes a collection of articles on waves in the marginal ice zone, spanning observations \citep{waseda2022observation},  physical modelling \citep{toffoli2022modelling}, numerical modelling \citep{perrie2022modelling}, a review of wave dissipation theory \citep{shen2022wave}, and a general overview \citep{thomson2022wave}.
Concise reviews of waves in the marginal ice zone appear in broader review articles on modelling sea ice \citep{golden2020modeling} and Southern Ocean dynamics \citep{bennetts2024closing}, and of waves in landfast ice in a review of that ice type \citep{fraser2023antarctic}.
There is currently no review of research on waves in ice shelves. 

In this chapter, we give a chronology of findings from field observations of waves in landfast ice, ice shelves and the marginal ice zone, from the contemporary perspective of the key research questions outlined above for each of the three ice types.
We follow this with an introduction to theoretical waves-in-ice models, with an emphasis on the close connections between the theories between the different ice types.
Thus, the chapter synthesises waves-in-ice knowledge derived from observations and the associated theories.
Physical models of waves in ice and numerical models that incorporate waves-in-ice theories are not covered.
Further, waves in ice due to local sources, such as those created by winds over the marginal ice zone, moving loads on landfast ice and icequakes in ice shelves, are considered out-of-scope.
Within these confines, the literature covered for waves in landfast ice and ice shelves is intended to be near comprehensive, but the large corpus of literature on waves in the marginal ice zone has meant that the studies reviewed are chosen to best illustrate the themes of the chapter. 

\section{Field observations} 

\subsection{Landfast ice}

Observations of waves in landfast ice were once considered to be more challenging than those in the marginal ice zone, as early attempts failed due to rapid breakup of the ice before the instruments were deployed and fully functioning \citep{squire1995ocean}.
The first field experiment to claim limited success was conducted in Newfoundland during 1977, in which ice-coupled waves propagating through landfast ice were recorded by a set of closely spaced devices and used to calculate the dispersion relation \citep{squire1977propagation}.
A far more extensive set of field measurements were collected as an opportunistic side experiment during MIZEX 1983, on $\approx{}1$\,m thick ``mushy'' landfast ice in the fjords of Svalbard, such that the ice was attached to land on all sides but at the ice edge \citep{squire1984theoretical}.
Two vertical accelerometers were deployed on the landfast ice cover, with one device $\approx{}5$\,m from the ice edge and the second device farther from the ice edge, and moved between different locations up to $400$\,m from the first device and recording for 0.5\,h on each deployment.
Significant wave heights up to 0.1\,m were measured, and cracks were observed to appear during the experiment up to $\approx{}$50\,m from the ice edge, which were attributed to wave-induced ice flexure.
The measurements showed ice-coupled wave energy attenuates by an order of magnitude or more over only a few hundred metres, such that short period wave components experience the strongest attenuation.
The components of the wave energy (density) spectrum were shown to display local extrema within the first few tens of metres from the ice edge, followed by an approximately exponential rate of decay away from the ice edge.
Exponential attenuation rates of wave energy, $\alpha$, were found by fitting exponential curves to the data points, with values on the order of $10^{-3}$ per metre at wave periods 6.8--8.5\,s and $10^{-2}$ at 5.8\,s.
  
More recent field observations of wave propagation through landfast ice have been made in both the Arctic and Southern oceans.
Similar to MIZEX~1983 \citep{squire1984theoretical}, one experiment was conducted in a fjord of Svalbard \citep{sutherland2016observations}.
The experiment was conducted continuously over three days during March~2015, using three triaxial accelerometers deployed on $0.5$--0.6\,m thick landfast ice.
One device was deployed $\approx{}100$\,m from the ice edge and the two remaining devices were deployed close to one another, around 50\,m farther onto the ice cover.
The dispersion relation was found to be gravity dominated for low frequencies (0.08--0.12\,Hz or 8.3--12.5\,s), transitioning to flexure dominated for higher frequencies, although the transition to high-frequency flexure dominance was lost after a day into the experiment, which was attributed to the appearance of cracks in the ice cover. 
Strong attenuation (up to 80\% between the sensors) was found only for frequencies $>0.15$\,Hz (or $<6.3$\,s) and before cracks appeared in the ice cover. 
There was evidence of counter propagating waves at high frequencies, which were inferred as wave scattering, although the scattering source was unknown. 
A subsequent experiment was conducted in a nearby Svalbard fjord and analysed alongside an experiment conducted on landfast ice north of Casey station in Antarctica, where the ice cover is not confined by sidewalls in a relatively narrow channel, as in the Svalbard fjords \citep{voermans2021wave}.
Two inertial motion units were deployed on the landfast ice to record waves in both experiments.
The Arctic experiment lasted two weeks and the ice was 0.3--0.4\,m thick, 
whereas the Antarctic experiment lasted three to four weeks during October~2020 and the ice was 1.1--1.3\,m thick.
The Arctic data showed attenuation rates, $\alpha$, that decrease from order $10^{-3}$ per metre at wave periods approximately 6\,s or below and then $10^{-4}$ per metre up to 15\,s. 
The Antarctic data were found to be unreliable for determining attenuation, and the results were scattered, although with magnitudes order $10^{-4}$ per metre, i.e., comparable with those from the Arctic.


\subsection{Marginal ice zone}

The first recordings of waves in the marginal ice zone were made in 1959--1960 (before the term marginal ice zone had been coined) on outbound and return voyages through the Antarctic ice pack in the Weddell Sea using a ship-borne recorder \citep{robin1963wave}.
The recordings were made for ten minutes every six hours and were accompanied by visual estimates of ice thickness, floe size and concentration.
The observations were the basis for a seminal study \citep{robin1963wave}, which identified many key processes that remain topics of research activity today, such as the relationship between wavelengths and floe lengths.
It took until the early 1970s for further observations to be reported \citep{wadhams1975airborne,wadhams1978wave}.
They were made in the Arctic, where the waves and ice were measured remotely by an airborne laser profiler \citep{wadhams1975airborne} and an echo sonar on a submarine \citep{wadhams1978wave}, which avoids contamination of the measurements by the ship. 
The accompanying studies were also seminal, as they introduced the paradigm that the frequency components of the wave energy spectrum attenuate exponentially with distance into the sea ice-covered ocean, and that the attenuation rate, $\alpha$, has a power-law relationship with frequency, $f$, of the form
\begin{equation}\label{eq:powerlaw_monomial}
\alpha =  c\,f^{n}.	
\end{equation}
The attenuation coefficients were found to be around order $10^{-4}$ per metre and the power-law exponent $n\approx{}2$--2.7.
They also highlighted the importance of concomitant observations of the ice cover properties, which they achieved using aerial photography and an infrared scanner \citep{wadhams1975airborne} 
	or inferred from the sonar and visual observations using a periscope \citep{wadhams1978wave}.
Greater attenuation was observed for more densely packed floe fields, which was attributed to frictional dissipation between floes \citep{wadhams1978wave}.

From 1978--1984, members of the Scott Polar Research Institute embarked on a series of experiments to measure wave evolution through Arctic marginal ice zones, using helicopters to move from floe to floe and deploying accelerometers on the floes.
The deployments on each floe were typically limited to tens of minutes and the analyses relied on the assumption that the incident field was statistically stationary over the experiment, such that the observations from multiple floes could be compared.
One experiment was conducted in the Bering Sea during spring 1979 in a marginal ice zone consisting of $\approx{}0.5$\,m-thick, ``mushy'' ice floes, with a 5\,km-wide, diffuse edge zone of $\approx{}10$\,m-diameter floes, and an interior zone of large floes (diameters $>100$\,m) starting 30\,km away from the ice edge, separated by a transition zone where the floe diameter steadily increased over distance up to 40\,m    \citep{squire1980direct}.
Another experiment was conducted in the Greenland Sea during MIZEX~1984 \citep{wadhams1985marginal}, which consisted of four runs at different locations and on different dates, where the reported floes were relatively thick (2--3\,m) and large (diameters 72--350\,m) and at a range of concentrations \citep{wadhams1986effect}.
The 1979 Bering Sea experiment \citep{squire1980direct} was revisited and compared against experiments in the Greenland Sea during September~1978, September~1979 and July~1983, and the Bering Sea during February~1983 \citep{wadhams1988attenuation}.
The findings support the concept of exponential attenuation  over distance of the wave energy spectral components, with the attenuation rate, $\alpha$, of order $10^{-4}$--$10^{-5}$~per~metre \citep{squire1980direct,wadhams1986effect,wadhams1988attenuation}.
The attenuation rate was found to increase linearly with ice thickness \citep{wadhams1988attenuation}.
Multi-axial devices were used during MIZEX~1984, such that the directional wave spectrum was retrieved \citep{wadhams1986effect}.
The observations showed that high frequencies (``wind seas'') broaden to become isotropic after $<5$\,km, whereas swell initially narrows before broadening and becomes isotropic after tens of kilometres \citep{wadhams1986effect}.
The broadening was attributed to wave scattering by floes, which competes with increased attenuation of directional components of the spectrum due to longer path lengths travelled to reach an observation location within the marginal ice zone.

By the early 1990s, analysis techniques for synthetic aperture radar (SAR) images had advanced to the point at which they could be used to study wave propagation in the marginal ice zone \citep{wadhams1991waves,liu1991wave,liu1991observation,larouche1992directional}.
SAR snapshots of the wave field over multiple-kilometre scales were used, and were processed to obtain wavenumber spectra.
Airborne SAR images obtained during LIMEX in March 1987 of compacted and rafted floes with $<20$\,m diameters in a brash ice matrix were combined with accelerometer and wave buoy measurements to compute the ice-coupled dispersion relation, as well as the attenuation rate of the peak spectral components \citep{liu1991wave} and provide evidence of refraction at the ice edge in the form of a wave energy cut-off beyond a critical incidence angle (similar to that found for landfast ice) \citep{liu1991observation}.
A subset of the LIMEX SAR data was re-assessed using a parametric spectral-density estimation technique \citep{larouche1992directional}, and used, for instance, to calculate the exponential attenuation rate for the low-frequency spectral components (periods $>12$\,s), which were found to be around half those calculated using accelerometer data, although noting the open-water dispersion relation was assumed in the calculations, and to provide evidence of wave refraction within the marginal ice zone, associated with a change in wavelength, which was attributed to a sharp change in ice concentration.
A mosaic of two satellite-borne SAR images of waves in pancake--frazil ice covers in the Chukchi Sea during October 1978 were used to derive the wavenumber spectra in subregions of the imagery \citep{wadhams1991waves}. 
The observed reduction in dominant wavelength with distance into the ice cover and slight refraction towards the normal with respect to the ice edge were compared with a theory (mass loading; \S\,\ref{sec:MIZtheory}), and combined with the theory to estimate the ice thickness.
More extensive studies of waves in frazil--pancake icefields were conducted following acquisition of additional SAR data, including two Arctic experiments (in April~1993 and March~1997) and one Antarctic experiment in July~1997 \citep{wadhams2002use}, and then a further Antarctic experiment in April~2000 \citep{wadhams2004sar}.
The studies provided evidence of a decrease in wavelength in the ice cover and refraction towards the normal direction.

From around the year 2000, technological advances allowed in situ observations of waves in the marginal ice zone to be made over weeks to months, with the data tranmitted via satellites.
An array of six buoys that each relayed (frequency) wave spectra every 3\,h (calculated from $\approx{}30$\,min timeseries of vertical accelerations), as well as the buoy locations, were deployed in the Weddell Sea marginal ice zone during advancing pancake--frazil ice condition in April 2000 \citep{doble2013wave,doble2015relating}.
Attenuation rates, $\alpha$, were calculated using observations from pairs of buoys
	over a 12-day period that covered an ice compression phase and a following re-expansion phase. 
A linear increase in the attenuation rate with increasing ice thickness was found during the compression phase, using model outputs for ice thickness \citep{doble2015relating}.
One of the buoys survived until October~2000, and its final two months of observations captured a large wave event that broke the ice cover, as inferred from satellite-derived ice concentrations, and showed a significant increase in wave energy reaching the buoy following the breakup event, indicating that waves propagate more easily through broken ice covers \citep{doble2013wave}. 

The wave buoy-array approach was extended during SIPEX~II, for which five bespoke buoys were deployed on the surfaces of ice floes in the East Antarctic marginal ice zone in September 2012 (which also relayed frequency spectra based on $\approx{}30$\,min timeseries every 3\,h) \citep{kohout2014storm,meylan2014situ}. 
The buoys provided concomitant observations of wave spectra at different distances into the marginal ice zone along a meridional transect, from 16\,km to 130\,km from the ice edge, for up to 39~days, although they lost their alignment as they drifted north-eastward.
The relative measurements of pairs of buoys indicated that the significant wave height, $H_{\text{s}}$ (a proxy for the integrated energy spectrum) attenuates at an exponential rate over distance for mild conditions ($H_{\text{s}}<3$\,m) and linearly for more energetic conditions ($H_{\text{s}}>3$\,m) \citep{kohout2014storm}, and that the exponential attenuation rate of the spectral components is related to frequency, such that 
\begin{equation}\label{eq:powerlaw_binomial}
	\alpha\approx{}a\,f^{2}+b\,f^{4},
\end{equation}
where $a=2.12\times10^{-3}$\,s$^{2}$\,m$^{-1}$ and $b=4.59\times10^{-2}$\,s$^{4}$\,m$^{-1}$ \citep{meylan2014situ}.

Wave buoy arrays have been deployed in the marginal ice zone during two subsequent field campaigns.
Six wave experiments were conducted in the Arctic marginal ice zone from October--November~2015 during the SeaState campaign, where the ice cover was dominated by pancake--frazil ice \citep{cheng2017calibrating,collins2018observations,montiel2018attenuation}.
Each wave experiment lasted hours to days, using up to seventeen wave buoys of three different types, where the buoys were recovered and reused for the subsequent experiments.
Directional wave spectra were estimated from the timeseries given by each buoy split into 30\,min segments. 
The buoy-pair approach (adapted to include wave direction) showed attenuation rates, $\alpha$, ranging over order $10^{-7}$--$10^{-2}$~per~metre \citep{cheng2017calibrating}. 
A single experiment involving a large wave event (up to $H_{\text{s}}\approx{}5$\,m measured) was analysed in detail, with the major findings being support for the transition from exponential to linear attenuation of the significant wave height at $H_{\text{s}}\approx 3$\,m, evidence of a similar (although less clearly defined) switch for the spectral components, and narrowing of the wave direction over distance \citep{montiel2018attenuation}.  
The same experiment was the focus of a study on wave dispersion, which found almost no deviation from open water dispersion for frequencies $<30$\,Hz and a slight increase in wavenumber relative to open water for higher frequencies \citep{collins2018observations}.

An extended version of the SIPEX~II waves-in-ice observations were made during the PIPERS campaign, in which fourteen wave buoys were deployed on floes along a meridional transect of the Ross~Sea marginal ice zone in autumn 2017 \citep{kohout2020observations,rogers2021estimates,montiel2022physical}.
Each buoy relayed the frequency wave spectra based on 11\,min timeseries, typically every 15\,mins, and operated for up to three months, creating the largest database of wave buoy observations to date and capturing large wave events, including significant wave heights $>9$\,m \citep{kohout2020observations}. 
In contrast to previous studies \citep{kohout2014storm,montiel2018attenuation}, the significant wave height was found to attenuate exponentially, even for large waves, although with indications for increases in the attenuation rate at higher concentrations and shorter periods \citep{kohout2020observations}.
The attenuation rates of the spectral components from the full dataset were investigated using the buoy-pair approach \citep{montiel2022physical}, and from a 24-day subset of the data by optimising the exponential attenuation rates of the spectral components in the WAVEWATCH~III model to match the observations \citep{rogers2021estimates}.
The studies support power-law relationships of either binomial form \eqref{eq:powerlaw_binomial} \citep{rogers2021estimates} or monomial form \eqref{eq:powerlaw_monomial} with exponent $n=3.5$--4 \citep{rogers2021estimates} or $n\approx{}3$ within a few tens of kilometres from the ice edge, decreasing to $n<2$ over 100\,km from the ice edge \citep{montiel2022physical}.
Both studies correlated changes in the attenuation rates with co-located variables (or ``physical drivers''), finding strong evidence that the attenuation rate increases with ice thickness and decrease with significant wave height \citep{rogers2021estimates}, and increases with opposing (southerly) winds \citep{montiel2022physical}. 

Some smaller scale buoy observations are also notable.
A single buoy was deployed in the winter Antarctic marginal ice zone during a cyclone that captured a significant wave height $>6$\,m at over 100\,km from the ice edge \citep{vichi2019effects,alberello2020drift}.
Another buoy captured observations for almost a year in the Antarctic ice pack, during which it detected a significant wave height $\approx{}0.1$\,m over 1000\,km from the ice edge \citep{nose2024observation}. 
Two drifting wave buoys operated in the western Antarctic marginal ice zone during winter 2018, with a third in the open ocean close to the ice edge observing the incident wave fields \citep{ardhuin2020ice}. 
The observations show wave fields at 200\,km from the ice edge with heights up to 1\,m and narrow directional distributions (spreads $<20^{\circ}$).

Over the past decade, in concert with the proliferation of in situ observations, there has been a resurgence in studies of waves in the marginal ice zone using remote sensing observations.
A technique was developed for measuring directional wavenumber spectra in the marginal ice zone using airborne scanning LIDAR \citep{sutherland2016airborne,sutherland2018airborne}, extending the previous single-point airborne laser profiling measurements \citep{wadhams1975airborne}.
The method was demonstrated for observations along a 60\,km transect of the Arctic marginal ice zone, taken from a aircraft over a 17\,min period in late April 2006, for a broken ice field in which the maximum floes sizes (captured from a camera on the aircraft) were $\approx{}50$\,m (less than half the dominant wavelength) \citep{sutherland2016airborne}.
The directional wave spectrum was calculated for each 4\,km segment of the flight, and used to show wave energy attenuation over distance, with a concomitant increase in peak wavelength and broadening of the directional spectrum \citep{sutherland2016airborne}.
Similar measurements were made from five aircraft flights during the SeaState campaign, and data from two of the flights were found to be usable for waves-in-ice analysis \citep{sutherland2018airborne}.  
For one flight, where the incoming wave field was near orthogonal to the ice edge, attenuation rates, $\alpha$, of order $10^{-4}$--$10^{-2}$~per~metre were calculated at up to 4\,km from the ice edge, which indicated a power-law frequency dependence \eqref{eq:powerlaw_monomial} with $n\approx{}7\,/\,4$.

Methods have been developed to estimate wave heights in the marginal ice zone from SAR imagery over transects hundreds of kilometres long, although limited to long waves (swell) and, thus, only applicable at sufficient distances from the ice edge for short-wave components (wind seas) to become negligible \citep{ardhuin2015estimates,ardhuin2017measuring,stopa2018wave,stopa2018strong}.
They have been applied to SAR data from Sentinel-1 satellites, including a set of images timed to coincide with the SeaState campaign that captured the large wave event during the campaign ($H_{\text{s}}>4$\,m), for which waves were detected $>100$\,km from the ice edge \citep{stopa2018wave}.
These observations showed exponential attenuation rates of significant waves heights of order $10^{-5}$~per~metre before a network of leads (visible in the SAR imagery), weakening to order $10^{-6}$~per~metre after the leads, which was hypothesised to result from the leads separating broken ice covers (before) from larger floes (after) \citep{stopa2018wave}.
More generally, Sentinel-1 satellites have been imaging Antarctic sea ice year-round since 2014, and over two thousand $20\times20$\,km$^{2}$ images with suitable wave and ice conditions were analysed to find significant wave height attenuation rates spanning three orders of magnitude, with a median of $3\times10^{-5}$ per metre \citep{stopa2018strong}.
A method to derive the 2D wave spectrum from SAR observations has also been developed and applied to images of the marginal ice zones of Svalbard and Greenland during March--April, 2021 \citep{huang2023wave}.
Attenuation rates of the resulting significant wave heights were order $10^{-5}$~per~metre across fourteen analysed transects that  included new, young and first-year ice, and the peak wave periods were 10--14\,s.

Laser altimeter measurements of vertical displacements of the ocean surface from the IceSat-2 satellite have been used to infer waves in the marginal ice zone \citep{horvat2020observing,brouwer2022altimetric,hell2024method}.
IceSat-2 has been operating since October~2018 and providing ``near instantaneous'' (ground speeds 7\,km\,s$^{-1}$) snapshots along transects of the Earth surface, including long stretches  (hundreds to thousands of kilometres) of the sea ice-covered oceans in both hemispheres  \citep{brouwer2022altimetric}.
They are inhibited by cloud cover, and the vertical displacements due to waves must be decoupled from those due to sea ice, such that only 10--15\% of the transects are usable \citep{brouwer2022altimetric}.
Wave attenuation has been identified from the displacements and used to define the marginal ice zone width \citep{horvat2020observing,brouwer2022altimetric}, and limited validation against wave buoy measurements has been attempted in terms of significant wave heights \citep{brouwer2022altimetric}.
A method to extract the wavenumber--direction wave spectra from altimeter measurements has been proposed and applied to a set of IceSat-2 transects \citep{hell2024method}.   

\subsection{Ice shelves}

As part of the International Geophysical Year program, 1957--1958, gravimeters that detect elevation changes were deployed on the upper surfaces of the Ross and Ronne--Filchner ice shelves, which are the two largest Antarctic ice shelves \citep{thiel1960gravimetric}.
The stations closest to the shelf fronts (2--5\,km away), where the shelves were $>200$\,m thick, recorded tidal signals overlaid by ``high frequency'' oscillations (15--50\,s periods), which were presumed to be ocean waves travelling through the shelves.
The high-frequency oscillations were greatly reduced at stations farther from the shelf fronts (10--15\,km away).
In February 1958, during a spell of extensive open water offshore from the Ross Ice Shelf, the high-frequency oscillations close to the shelf front became so large that they exceeded the threshold of the gravimeter.
Ice shelf oscillations attributed to ocean waves were also measured by gravimeters on the Ross Ice Shelf during the 1970s \citep{williams1981flexural}. 
A particular experiment used concurrent measurements from three stations at the southern end of the shelf, where the ice is 300--600\,m thick, which provided evidence that ocean waves in the ice shelf manifest as flexural waves with speeds 50--65\,m\,s$^{-1}$, although neglecting possible dispersive effects.
Evidence was also found that short period waves attenuate over distance, and of resonances around wave periods of 17\,s and 45\,s.

In contrast to the early observations made on giant ice shelves, strain gauges were deployed on the surface of the Erebus Ice Tongue in the 1980s, for which the floating part of the tongue is only $\approx{}10$\,km long, 0.5--2\,km wide, and from 50\,m thick at its snout (the seaward tip) to 300\,m at its grounding line \citep{robinson1992travelling,squire1994observations}.
The strain gauges provided measurements from November~1984 to November~1989, which missed a large calving event by only months \citep{robinson1990calving}. 
Maximum strains of $3\times10^{-7}$ were measured during a storm event \citep{robinson1992travelling}.
Evidence was found of waves travelling along the ice tongue, from the snout to the grounding line, with celerity $\approx{}70$\,m\,s$^{-1}$ and wave period $\approx{}50$\,s, and were attributed to infragravity waves, which were a recently discovered concept \citep{robinson1992travelling}.
Analysis of strain measurements on the surrounding sea ice over a few days in November~1989 also showed a 50\,s peak, and with greater energy density than on the ice tongue \citep{squire1994observations}.
  
A single broadband seismic station, consisting of one vertical and two horizontal seismometers,  was deployed on the Ross Ice Shelf from November~2004 to November~2006, close to an anticipated calving site known as the Nascent Iceberg Rift \citep{macayeal2006transoceanic,cathles2009seismic,bromirski2010transoceanic,bromirski2012response}. 
The seismometers were powered by sunlight, such that they recorded for 340 days outside of winter over the two-year deployment.
Relatively large-motion swell events ($\approx{}7$--40\,s periods) and infragravity wave events (50--250\,s) that originated from northern hemisphere storms were detected in spectrograms as slanting bands (caused by dispersion in the arrival time of the wave groups) \citep{macayeal2006transoceanic,cathles2009seismic,bromirski2010transoceanic}.
Swell created amplitudes up to 30\,mm \citep{cathles2009seismic} and infragravity waves up to approximately three times greater, which was attributed to amplification by shoaling being more significant for longer waves \citep{bromirski2010transoceanic}.

The Ross Ice Shelf observations were extended to a 34-station seismic array from November~2014 to November~2016, where the stations were arranged into two linear transects that were approximately parallel and orthogonal to the shelf front, and with a dense subarray at the intersection of the transects \citep{bromirski2015ross,bromirski2017tsunami,chen2018ocean,chen2019ross}.
The stations were powered by a combination of solar panels and lithium batteries, so that they could operate throughout the year. 
During austral summer, ten to twenty large swell events per month (10--30\,s wave periods) were detected, reaching vertical amplitudes up to 4\,mm at the stations closest to the shelf front, but showing significant attenuation away from the shelf front \citep{chen2018ocean}.
The horizontal amplitudes were smaller than the vertical amplitudes but attenuated more weakly away from the shelf front \citep{chen2018ocean}.
A large infragravity wave event (dominant energy in the 50--300\,s wave-period band) was detected in May~2015, where the vertical displacements towards the shelf front were up to almost 10\,mm but attenuating to $\approx{}1$\,mm at 350\,km from the shelf front \citep{bromirski2017tsunami}.
Towards the shelf front, infragravity waves create near continuous vertical displacement of $\approx{}2$\,mm amplitude \citep{chen2019ross}, presumed to be due to infragravity waves bound to swell, as opposed to free infragravity waves that leak away from distant coastlines to create the large events \citep{bromirski2015ross}.  
A tsunami event was also captured, with dominant energy in the very-long period regime (300--1000\,s wave-period band), which created vertical amplitudes over 10\,mm without appreciable attenuation away from the shelf front, and were amplified at the station above a seabed protrusion \citep{bromirski2017tsunami}.
The vertical displacements from one of the stations nearest the shelf front was analysed alongside observations of incoming waves from a nearby hydrophone mounted to the seabed just north of the shelf front, which indicated vertical ice shelf displacements relative to ocean wave displacements increase with wave period from order $10^{-2}$ at 30\,s to just below unity at 100\,s, and are relatively insensitive to wave period above 100\,s \citep{chen2019ross}. 
Beamforming was also used to generate dispersion curves from the coherent wave signals over the dense subarray, and gave evidence of flexural-gravity waves for wave periods $<50$\,s from both the vertical and horizontal motions, and much faster extensional Lamb waves for 10--50\,s wave periods from the horizontal motions \citep{chen2018ocean}.
In contrast, flexural-gravity waves extended into the swell regime from observation by a five-station array on the Pine Island Glacier during 2012--2013, which was attributed to the stations being closer to the shelf front, such that the swell had not attenuated \citep{chen2018ocean}.

\section{Theoretical models}

\subsection{Waves in landfast ice}\label{sec:theory:landfast}

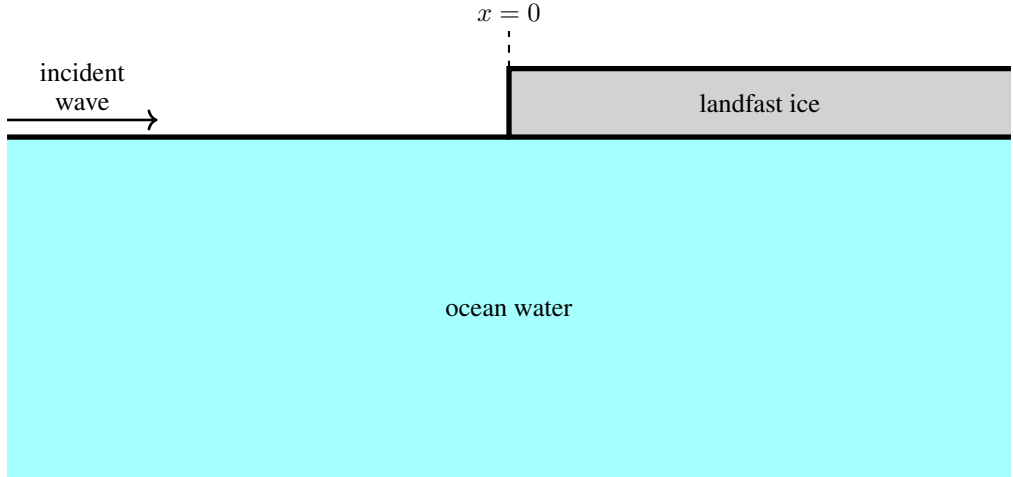
\begin{figure}[h!]
        \centering
        \setlength{\figurewidth}{0.85\textwidth} 
        \setlength{\figureheight}{0.4\textwidth} 
%
\definecolor{mycolor1}{rgb}{0.00000,1.00000,1.00000}%

\def\labelsA{1} 
\def\labelsB{0} 
\def\labelsC{0} 
\def\labelsD{1} 

\def\L{3000} 
\def\H{200}  
\def\d{0}   
\def\f{40}   

\begin{tikzpicture}

\begin{axis}[%
width=0.951\figurewidth,
height=\figureheight,
at={(0\figurewidth,0\figureheight)},
scale only axis,
xmin=-\L,
xmax=\L,
xtick={},
xlabel style={font=\color{white!15!black}},
xticklabels={},
xlabel={$x$},
ymin=-210,
ymax=80,
ytick={-200,0},
yticklabels={$-h_{0}$,0},
ylabel style={font=\color{white!15!black}},
ylabel={$z$},
axis background/.style={fill=white},
hide axis
]

\coordinate (origin) at (axis cs:0,0);
\coordinate (e1) at (axis cs:1,0);
\coordinate (e2) at (axis cs:0,1);

\coordinate (A) at (axis cs:-\L,-\H);
\coordinate (B) at (axis cs:0,-\H);
\coordinate (C) at (axis cs:\L,-\H);
\coordinate (D) at (axis cs:\L,-\d);
\coordinate (E) at (axis cs:0,-\d);
\coordinate (F) at (axis cs:-\L,0);
\coordinate (G) at (axis cs:0,\f);
\coordinate (H) at (axis cs:\L,\f);
\coordinate (I) at (axis cs:-\L,-\d);

\fill[mycolor1!35, domain=0:4000, variable=\x, samples=200] (A) -- (B) -- (C) -- (D) -- (E) -- (origin) -- (F) --cycle;

\fill[gray!35, line width=2.0pt, domain=0:4000, variable=\x, samples=200] (D) -- (E) -- (G) -- (H) -- cycle;

\draw[black, line width=2.0pt] (D) -- (E) -- (G) -- (H);  
\draw[black, line width=2.0pt, domain=0:4000, variable=\x, samples=200] (F) -- (origin);

\draw[->, black, line width=1.0pt] (axis cs: -\L,0.25*\f)  -- node[above,align=center]{incident \\ wave} (axis cs: -0.7*\L,0.25*\f);

%

\if\labelsA1
\node[align=center,xshift=0] at ($(E)!0.5!(H)$) {landfast ice};
\node[align=center] at ( $ (B)!0.5!(E) $ ) {ocean water}; 
\fi
\if\labelsB1
\draw[latex-latex,black,thick] ($(B)!0.9!(A)$) -- node[right] {$H$} ($(origin)!0.9!(F)$);
\draw[latex-latex,black,thick] ($(E)!0.025!(I)$) -- node[left] {$d$} ($(origin)!0.025!(F)$);
\draw[latex-latex,black,thick] ($(E)!0.9!(D)$) -- node[left] {$h$} ($(G)!0.9!(H)$);
\fi
\if\labelsC1
\draw[-latex,thick,matlabred] (origin) -- ++(0:1.5cm) node[right] {$x$};
\draw[-latex,thick,matlabred] (origin) -- ++(90:1.5cm) node[above] {$z$};
\fi
\if\labelsD1
\draw[dashed,thick,black] (D) -- ++(0:0.5cm) node[right] {$z=0$};
\draw[dashed,thick,black] (G) -- ++(90:0.5cm) node[above] {$x=0$};
\fi

\end{axis}

\end{tikzpicture}%
        \caption{Schematic (not to scale) of the {equilibrium} geometry for the standard theoretical model of ocean wave interactions with landfast ice.}
        \label{fig:schematic-landfast}
\end{figure}

The standard theoretical model of ocean waves propagating into and through landfast ice treats the ice as a thin elastic Kirchhoff plate (or Euler--Bernoulli beam), floating on water that is modelled using potential-flow theory, i.e., the water is inviscid, incompressible and undergoes irrotational motions.
Linear conditions are imposed, under the assumption that the amplitudes of motion are much smaller than the characteristic wavelengths (i.e., the waves have small steepness). 
The canonical problem involves a water domain of infinite horizontal extent and ice of uniform properties and covering half of the water surface, so that the other half is open water, from which incident wave forcing is prescribed  (Fig.~\ref{fig:schematic-landfast}).
The water domain is typically assumed to be bounded below by a flat impermeable seabed at a finite depth, $H$ \citep{fox1990reflection,fox1994oblique,chung2001calculation}.

Let $t$ denote time, and the Cartesian coordinate system $(x,y,z)$ denote locations in the water domain, where $(x,y)$ denotes the horizontal location and $z$ the vertical location.
Without loss of generality, the ice edge is aligned along the $y$-axis ($x=0$) and the origin of the vertical coordinate is located at the undisturbed water surface (Fig.~\ref{fig:schematic-landfast}).
For convenience in solving the problem, it is common to assume the ice has no draught, so that its lower surface occupies the plane $z=0$ for $x>0$. 
Due to the thin-plate assumption, the flexural ice motion is determined solely from the vertical displacements of its lower surface, $\zeta(x,y,t)$ ($x>0$).
The function $\zeta$ extends to the open water region ($x<0$) to denote the vertical displacements of the free surface.
The water velocity field is defined as the gradient of a scalar function, $\Phi(x,y,z,t)$, known as a velocity potential, which satisfies Laplace's equation throughout the water domain.

It is convenient to map the problem from the time domain to the frequency domain (implicitly using a Fourier transform), and consider a time-harmonic problem at an arbitrary angular frequency, $\omega=2\,\pi\,f$.
Thus, the unknown surface displacement and velocity potential functions are, respectively,
\begin{equation}
\zeta(x,y,t)=\text{Re}\{A_{\text{inc}}\,\eta(x,y)\,\e^{-\ci\,\omega\,t}\}
\mathand
\Phi(x,y,z,t)=\text{Re}\Big\{\frac{g\,A_{\text{inc}}}{\ci\,\omega}\phi(x,y,z)\,\e^{-\ci\,\omega\,t}\Big\}	,
\end{equation}
where $A_{\text{inc}}$ is the incident wave amplitude, $\ci=\sqrt{-1}$ is the imaginary unit, $g=9.81$\,m\,s$^{-2}$ is the constant of gravitational acceleration, and $\eta$ and $\phi$ are complex-valued functions.
In the ice-covered region ($x>0$), they are coupled at their common interface by the conditions
\begin{subequations}\label{eq:water_ice_couple}
\begin{equation}
 F\,\nabla_{\perp}^{4}\eta - \omega^{2}\,m\,\eta = \rho_{w}\,g\{\phi - \eta\}
 \mathand
  \D{z}{\phi}=\frac{\omega^{2}}{g}\,\eta
  \qquad
  (z=0),
  \tag{\theequation a,b}
\end{equation}
\end{subequations}
where $\nabla_{\perp}\equiv (\partial\,/\partial{}x,\partial\,/\partial{}y)$, $F$ is the flexural rigidity of the ice, $m$ is its mass per unit area, and $\rho_{w}$ is the water denity.
Eq.~(\ref{eq:water_ice_couple}a) is a dynamic condition that equates pressure exerted by the ice from thin-plate theory (on the left-hand side) with the water pressure from linearised Bernoulli theory (on the right). 
Eq.~(\ref{eq:water_ice_couple}b) is a kinematic condition that sets the vertical velocity of the particles at the water surface (left-hand side) to be equal to the vertical velocity of the water surface.
The coupling conditions assume the lower surface of the ice and the surface of the water below are in contact at all points where $x>0$, and at all times during the motion.
Free-edge conditions are also applied to the ice edge, such that
\begin{equation}\label{eqs:free-edge}
	\DD{x}{\eta} + \nu\,\DD{y}{\eta} = 0
	\mathand
	\D{x}{}\Bigg\{\DD{x}{\eta} + (2-\nu)\,\DD{y}{\eta}\Bigg\} = 0
	\quad (x=0),
\end{equation}  
where $\nu$ is Poisson's ratio,
which represent vanishing of bending moment and shear stress, respectively.
In the open water region ($x<0$), the coupling conditions are
\begin{subequations}\label{eq:water_air_couple}
\begin{equation}
	\eta=\phi 
	\mathand
    \D{z}{\phi}=\frac{\omega^{2}}{g}\,\eta
	\quad
    (z=0),
    \tag{\theequation a,b}
\end{equation}
\end{subequations}
where the dynamic condition (\ref{eq:water_air_couple}a) is a degenerate version of (\ref{eq:water_ice_couple}a), 
and the kinematic condition (\ref{eq:water_air_couple}b) is unchanged from (\ref{eq:water_ice_couple}b).

The ice displacement can be eliminated from the coupling conditions (\ref{eq:water_ice_couple}a--b) and (\ref{eq:water_air_couple}a--b) to leave
\begin{subequations}
\begin{align}
	 F\,\nabla_{\perp}^{4}\D{z}{\phi} - \omega^{2}\,m\,\D{z}{\phi} & = \rho_{w}\,g\Bigg\{\frac{\omega^{2}}{g}\,\phi - \D{z}{\phi}\Bigg\} \quad & (x>0,z=0),
	 \\[4pt]
	 \mathandR
	 \D{z}{\phi} & =\frac{\omega^{2}}{g}\,\phi \quad & (x<0,z=0).
\end{align}
\end{subequations}
The free-edge conditions (\ref{eqs:free-edge}a--b) can also be expressed in terms of the velocity potential, as
\begin{equation}\label{eqs:free-edge-phi}
	\DD{x}{}\D{z}{\phi} + \nu\,\DD{y}{}\D{z}{\phi}  = 0
	\mathand
	\D{x}{}\Bigg\{\DD{x}{}\D{z}{\phi}  + (2-\nu)\,\DD{y}{}\D{z}{\phi} \Bigg\} = 0
	\quad (x=0,z=0).
\end{equation}  
These act as boundary conditions for Laplace's equation on the linearised water domain, 
\begin{equation}
\nabla^{2}\phi = 0
\quad
(x,y\in\mathbb{R},-H<z<0),
\mathwhere
\nabla\equiv (\partial\,/\partial{}x,\partial\,/\partial{}y,\partial\,/\partial{}z).
\end{equation}
The seabed condition is
\begin{equation}
\D{z}{\phi}=0
\quad (z=-H),	
\end{equation}
which enforces no normal flow, assuming the seabed is impermeable.

Seeking separation solutions, $\phi(x,y,z)=X(x,y)\,Z(z)$, leads to the vertical modes
\begin{subequations}\label{eq:water_ivertical_modes}
\begin{align}
	Z(z) & = \cosh\{k\,(z+h)\}
	\quad\text{in the open water } (x<0),
	\\[4pt]
	\mathandR Z(z) & = \cosh\{\kappa\,(z+h)\}
	\quad\text{in the ice-covered water } (x>0),
\end{align}
\end{subequations}
with associated horizontal modes
\begin{subequations}\label{eq:horizontal_modes}
\begin{equation}
X(x,y) = \e^{\pm\ci\,(k_{x}\,x+k_{y}\,y)}
\quad (x<0)
\mathand
X(x,y) = \e^{\pm\ci\,(\kappa_{x}\,x+\kappa_{y}\,y)}
\quad (x>0),
\tag{\theequation a,b}
\end{equation}
\end{subequations}
such that the wavevectors $\mathbf{k}=(k_{x},k_{y})$ and $\boldsymbol{\kappa}=(\kappa_{x},\kappa_{y})$ have magnitudes $\vert\mathbf{k}\vert=k$ and $\vert\boldsymbol{\kappa}\vert=\kappa$.
The wavenumbers $k$ and $\kappa$ are the roots of dispersion relations, respectively,
\begin{subequations}\label{eq:dispersion_relations}
\begin{equation}
k\,g\,\tanh(k\,H) = \omega^{2}
\mathand
\{F\,\kappa^{4} + \rho_{w}\,g - m\,\omega^{2}\}\,	\kappa\,\tanh(\kappa\,H) = \rho_{w}\,\omega^{2}.
\tag{\theequation a,b}
\end{equation}
\end{subequations}
Eq.~(\ref{eq:dispersion_relations}a) is the classical open water dispersion relation \citep{linton2001handbook}.
It has roots $k=\pm{}k_{0},\pm{}k_{1}, \ldots$, where $k_{0}\in\mathbb{R}_{+}$ supports propagating surface gravity waves, and $k_{n}\in\ci\mathbb{R}_{+}$ ($n=1,2,\ldots$), ordered such that $\vert{}k_{1}\vert<\vert{}k_{2}\vert<\ldots$, support so-called evanescent wave modes that decay exponentially away from a scattering source, such as an ice edge. 

Eq.~(\ref{eq:dispersion_relations}b) is the dispersion relation for a thin floating elastic plate, which has roots $\kappa=\pm\kappa_{-2}, \pm\kappa_{-1}, \pm\kappa_{0}, \pm\kappa_{1}, \ldots$.
Similar to the open water dispersion relation, $\kappa_{0}\in\mathbb{R}_{+}$, which supports propagating ice-coupled waves known as flexural-gravity waves, and $\kappa_{n}\in\ci\mathbb{R}_{+}$ ($n=1,2,\ldots$), such that $\vert{}\kappa_{1}\vert<\vert{}\kappa_{2}\vert<\ldots$, support evanescent wave modes.
Flexural-gravity waves are shorter than gravity waves for short periods (i.e., $\kappa_{0}>k_{0}$), for which mass loading dominates, and longer for long periods ($\kappa_{0}>k_{0}$) for which ice flexure dominates \citep{squire1977propagation,voermans2021wave}, although certain observations suggest the effect of the ice cover becomes negligible when cracks appear in the ice \citep{sutherland2016observations}.
The wavenumbers $\kappa_{-j}$ ($j=1,2$) have no analogue in open water.
They are typically complex valued, such that $\kappa_{-1}$ is in the first quadrant of the complex plane and $\kappa_{-2}=-\overline{\kappa_{-1}}$, for which they support so-called damped-propagating waves \citep{squire1995ocean}.
However, they can also appear on the imaginary axis (similar to evanescent wavenumbers), and in these situations $\kappa_{-1}$ and $\kappa_{-2}$ no longer maintain their skew-conjugate relationship \citep{williams2006reflections,bennetts2007wave}.

Let the incident wave from the open ocean, $\phi_{\text{inc}}$, be propagating towards the ice cover (in the positive $x$-direction) at an angle $\psi\in[0,\pi\,/\,2)$ to the positive $x$-axis, so that
\begin{equation}
\phi_{\text{inc}}=\e^{\ci\,k_{0}\,\{\cos(\phi)\,x + \sin(\psi)\,y\}} \,\frac{\cosh\{k_{0}\,(z+H)\}}{\cosh(k_{0}\,H)}.
\end{equation}
Uniformity of the geometry in the $y$-direction implies that the $y$-dependence of the incident wave can be enforced on the full solution, so that
\begin{equation}
	\phi(x,y,z)=\varphi(x,z)\,\e^{\ci\,k_{0}\,\sin(\psi)\,y}.
\end{equation}
Applying this restriction to the $y$-components of the wavevectors $\mathbf{k}$ and $\boldsymbol{\kappa}$, i.e., $k_{y}=\kappa_{y}=k_{0}\,\cos(\psi)$, means the $x$-components are 
\begin{subequations}\label{eq:wavenos_snells}
\begin{equation}
k_{x,n}^{2} = k_{n}^{2} - k_{0}^{2}\,\cos^{2}(\psi)
\quad
(n=0,1,\ldots)
\mathand
\kappa_{x,n}^{2} = \kappa_{n}^{2} - k_{0}^{2}\,\cos^{2}(\psi) 	
\quad
(n=-2,-1,0,1,\ldots).
\tag{\theequation a,b}
\end{equation}
\end{subequations} 
Eq.~(\ref{eq:wavenos_snells}b) results in two generic cases, with one case when $\kappa_{0}\geq{}k_{0}$, for which $\kappa_{x}\in\mathbb{R}_{+}$ for all incident angles, so that a propagating wave exists in the ice-covered region, and the case where $\kappa_{0}<k_{0}$, for which there is a critical angle $\psi_{\textrm{crit}}=\arccos(\kappa_{0}\,/\,k_{0})$ that divides existence of a propagating wave mode ($\kappa_{x}\in\mathbb{R}_{+}$) for $\psi<\psi_{\textrm{crit}}$ from decaying modes only ($\kappa_{x}\in\ci\,\mathbb{R}_{+}$) for $\psi>\psi_{\textrm{crit}}$.

The velocity potential, $\varphi$, is expressed as a linear superposition of wave modes defined by the dispersion relation in the relevant region (open or ice-covered water).
In the open water region ($x<0$), the wave field is the sum of the incident wave field, a leftward propagating reflected wave (amplitude $r_{0}^{\text{(la)}}$) and an infinite sum of evanescent waves that decay away from the ice edge (amplitudes $r_{n}^{\text{(la)}}$ for $n=1,2,\ldots$), so that
\begin{equation}\label{eq:emm_left}
\varphi(x,z)=\e^{\ci\,k_{x,0}\,x}\,\frac{\cosh\{k_{0}\,(z+H)\}}{\cosh(k_{0}\,H)} 
+ \sum_{n=0}^{\infty}r_{n}^{\text{(la)}}\,\e^{-\ci\,k_{x,n}\,x}\,\frac{\cosh\{k_{n}\,(z+H)\}}{\cosh(k_{n}\,H)}
\quad (x<0).
\end{equation}
In the ice-covered region ($x>0$), the wave field is a sum of a rightward propagating flexural-gravity wave (below the critical angle; amplitude $\tau_{0}^{\text{(la)}}$),
    the damped propagating waves  (amplitudes $\tau_{-n}^{\text{(la)}}$ for $n=1,2$) and an infinite number of evanescent waves (amplitudes $\tau_{n}^{\text{(la)}}$ for $n=1,2,\ldots$ below the critical angle and $n=0,1,\ldots$ above it) that decay away from the ice edge, so that 
\begin{equation}\label{eq:emm_right}
\varphi(x,z)= \sum_{n=-2}^{\infty}\tau_{n}^{\text{(la)}}\,\e^{\ci\,\kappa_{x,n}\,x}\,\frac{\cosh\{\kappa_{n}\,(z+H)\}}{\cosh(\kappa_{n}\,H)}
\quad (x>0). 
\end{equation}
The amplitudes associated with the two damped-propagating modes can be viewed as providing the degrees of freedom to satisfy the free-edge conditions (\ref{eqs:free-edge-phi}a--b).
The amplitudes of the propagating and evanescent wave modes in (\ref{eq:emm_left}--\ref{eq:emm_right}) then give the freedom to enforce continuity of pressure and horizontal velocity in the water column below the ice edge ($x=0$, $-H<z<0$).

For normal incidence ($\psi=0$), activation of all the wave modes ($n=-2,-1,0,1,\ldots$) results in the only non-zero component of the ice strain (normal to the ice edge) increasing from zero at the ice edge to a constant amplitude once the damped-propagating and evanescent wave modes have died out away from the ice edge, with a maximum strain occurring inbetween for shorter wave periods   \citep{fox1994oblique}.
The behaviour is less simple for non-normal incidence and below the critical angle, noting that the strain tensor has more than one non-zero component in this case \citep{fox1994oblique}.
For incidence at the critical angle and above it, the wave field can be expressed as a wave that travels parallel to the ice edge and decays away from it \citep{squire1984critical}. 

The model outlined above does not produce the observed attenuation of waves through the landfast-ice-covered ocean \citep{squire1984theoretical,sutherland2016observations,voermans2021wave}.
The standard model can be modified to include viscoelastic damping by using a complex-valued flexural rigidity, which was found to give reasonable agreement with observations of relatively large attenuation rates for near-melting landfast ice \citep{squire1984theoretical}.
An alternative modification incorporates damping using a term proportional to the ice displacement velocity \citep{robinson1990modal}, resulting in the dispersion relation 
\begin{equation}\label{eq:dr_RP} 
\{F\,\kappa^{4} + \rho_{w}\,g - m\,\omega^{2} - \ci\,\omega\,\gamma\}\,	\kappa\,\tanh(\kappa\,H) = \rho_{w}\,\omega^{2},
\end{equation}
where the damping parameter $\gamma=10$\,kPa\,m$^{-1}$ was found to predict attenuation in good agreement with one set of observations \citep{squire1992ice,squire1993breakup}.
A range of wave damping models, broadly divided into those in which the damping occurs in the ice layer and those in which it occurs in the underlying water  (in general, models not originally proposed for landfast ice), were compared against observations, with support for viscoelastic damping at shorter periods, and damping due to under ice turbulence and friction for longer periods \citep{voermans2021wave}.

\subsection{Waves in the marginal ice zone}\label{sec:MIZtheory}

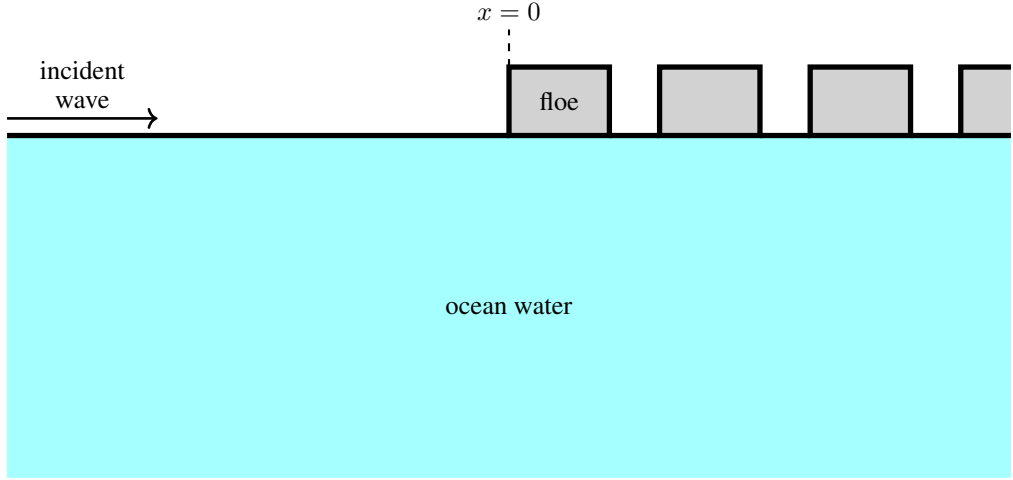
\begin{figure}[h!]
        \centering
        \setlength{\figurewidth}{0.85\textwidth} 
        \setlength{\figureheight}{0.4\textwidth} 
%
\definecolor{mycolor1}{rgb}{0.00000,1.00000,1.00000}%

\def\labelsA{1} 
\def\labelsB{0} 
\def\labelsC{0} 
\def\labelsD{1} 

\def\L{3000} 
\def\H{200}  
\def\d{0}   
\def\f{40}   

\begin{tikzpicture}

\begin{axis}[%
width=0.951\figurewidth,
height=\figureheight,
at={(0\figurewidth,0\figureheight)},
scale only axis,
xmin=-\L,
xmax=\L,
xtick={},
xlabel style={font=\color{white!15!black}},
xticklabels={},
xlabel={$x$},
ymin=-210,
ymax=80,
ytick={-200,0},
yticklabels={$-h_{0}$,0},
ylabel style={font=\color{white!15!black}},
ylabel={$z$},
axis background/.style={fill=white},
hide axis
]

\coordinate (origin) at (axis cs:0,0);
\coordinate (e1) at (axis cs:1,0);
\coordinate (e2) at (axis cs:0,1);

\coordinate (A) at (axis cs:-\L,-\H);
\coordinate (B) at (axis cs:0,-\H);
\coordinate (C) at (axis cs:\L,-\H);
\coordinate (D) at (axis cs:\L,-\d);
\coordinate (E) at (axis cs:0,-\d);
\coordinate (F) at (axis cs:-\L,0);
\coordinate (G) at (axis cs:0,\f);
\coordinate (H) at (axis cs:\L,\f);
\coordinate (I) at (axis cs:-\L,-\d);

\fill[mycolor1!35, domain=0:4000, variable=\x, samples=200] (A) -- (B) -- (C) -- (D) -- (E) -- (origin) -- (F) --cycle;

\fill[gray!35, line width=2.0pt] (axis cs:0.3*0*\L,-\d) -- (axis cs:0.3*0*\L,\f) -- (axis cs:0.3*0*\L+0.2*\L,\f) -- (axis cs:0.3*0*\L+0.2*\L,-\d)  -- cycle;
\fill[gray!35, line width=2.0pt] (axis cs:0.3*1*\L,-\d) -- (axis cs:0.3*1*\L,\f) -- (axis cs:0.3*1*\L+0.2*\L,\f) -- (axis cs:0.3*1*\L+0.2*\L,-\d)  -- cycle;
\fill[gray!35, line width=2.0pt] (axis cs:0.3*2*\L,-\d) -- (axis cs:0.3*2*\L,\f) -- (axis cs:0.3*2*\L+0.2*\L,\f) -- (axis cs:0.3*2*\L+0.2*\L,-\d)  -- cycle;
\fill[gray!35, line width=2.0pt] (axis cs:0.3*3*\L,-\d) -- (axis cs:0.3*3*\L,\f) -- (axis cs:0.3*3*\L+0.1*\L,\f) -- (axis cs:0.3*3*\L+0.1*\L,-\d)  -- cycle;

\draw[black, line width=2.0pt] (axis cs:0.3*0*\L,-\d) -- (axis cs:0.3*0*\L,\f) -- (axis cs:0.3*0*\L+0.2*\L,\f) -- (axis cs:0.3*0*\L+0.2*\L,-\d) -- cycle;  
\draw[black, line width=2.0pt] (axis cs:0.3*1*\L,-\d) -- (axis cs:0.3*1*\L,\f) -- (axis cs:0.3*1*\L+0.2*\L,\f) -- (axis cs:0.3*1*\L+0.2*\L,-\d) -- cycle; 
\draw[black, line width=2.0pt] (axis cs:0.3*2*\L,-\d) -- (axis cs:0.3*2*\L,\f) -- (axis cs:0.3*2*\L+0.2*\L,\f) -- (axis cs:0.3*2*\L+0.2*\L,-\d) -- cycle; 
\draw[black, line width=2.0pt] (axis cs:0.3*3*\L+0.1*\L,-\d) -- (axis cs:0.3*3*\L,-\d) -- (axis cs:0.3*3*\L,\f) --  (axis cs:0.3*3*\L+0.1*\L,\f); 
\draw[black, line width=2.0pt, domain=0:4000, variable=\x, samples=200] (axis cs:0.3*0*\L+0.2*\L,-\d) -- (axis cs:0.3*1*\L,-\d);
\draw[black, line width=2.0pt, domain=0:4000, variable=\x, samples=200] (axis cs:0.3*1*\L+0.2*\L,-\d) -- (axis cs:0.3*2*\L,-\d);
\draw[black, line width=2.0pt, domain=0:4000, variable=\x, samples=200] (axis cs:0.3*2*\L+0.2*\L,-\d) -- (axis cs:0.3*3*\L,-\d);
\draw[black, line width=2.0pt, domain=0:4000, variable=\x, samples=200] (F) -- (origin);

\draw[->, black, line width=1.0pt] (axis cs: -\L,0.25*\f)  -- node[above,align=center]{incident \\ wave} (axis cs: -0.7*\L,0.25*\f);

\if\labelsA1
\node[align=center,xshift=0] at (axis cs: 0.1*\L,0.5*\f) {floe};
\node[align=center] at ( $ (B)!0.5!(E) $ ) {ocean water}; 
\fi
\if\labelsB1
\draw[latex-latex,black,thick] ($(B)!0.9!(A)$) -- node[right] {$H$} ($(origin)!0.9!(F)$);
\draw[latex-latex,black,thick] ($(E)!0.025!(I)$) -- node[left] {$d$} ($(origin)!0.025!(F)$);
\draw[latex-latex,black,thick] ($(E)!0.9!(D)$) -- node[left] {$h$} ($(G)!0.9!(H)$);
\fi
\if\labelsC1
\draw[-latex,thick,matlabred] (origin) -- ++(0:1.5cm) node[right] {$x$};
\draw[-latex,thick,matlabred] (origin) -- ++(90:1.5cm) node[above] {$z$};
\fi
\if\labelsD1
\draw[dashed,thick,black] (D) -- ++(0:0.5cm) node[right] {$z=0$};
\draw[dashed,thick,black] (G) -- ++(90:0.5cm) node[above] {$x=0$};
\fi

\end{axis}

\end{tikzpicture}%
        \caption{Schematic (not to scale) of the {equilibrium} geometry for a standard theoretical model of ocean wave propagation through the marginal ice zone.}
        \label{fig:schematic-miz}
\end{figure}

In the classical model of waves in the marginal ice zone, the ice cover is treated as an array of  floes separated by open water.
The two-dimensional version of the model appears similar to the standard model for waves in landfast ice, except that the ice-covered region ($x>0$) consists of multiple elastic plates of finite length (Fig.~\ref{fig:schematic-miz}).
Exponential wave attenuation over distance results, without damping, from an accumulation of scattering events by the individual floes, in which energy is reflected back towards the open ocean rather than being dissipated.

Consider the individual-floe version of the model, where the floe occupies the interval $x\in(0,\ell)$. 
On the left-hand side of the floe, the wave field is the sum of incident and reflected waves, plus evanescent waves generated at the floe edge, such that
\begin{equation}
\varphi(x,z)=\e^{\ci\,k_{x,0}\,x}\,\frac{\cosh\{k_{0}\,(z+H)\}}{\cosh(k_{0}\,H)} 
+ \sum_{n=0}^{\infty}r_{n}^{\text{(fl)}}\,\e^{-\ci\,k_{x,n}\,x}\,\frac{\cosh\{k_{n}\,(z+H)\}}{\cosh(k_{n}\,H)}
\quad (x<0),
\end{equation}
which is identical in form to Eq.~(\ref{eq:emm_left}).
On its right-hand side, the wave field is a transmitted wave plus evanescent waves, such that
\begin{equation}
\varphi(x,z)= \sum_{n=0}^{\infty}t_{n}^{\text{(fl)}}\,\e^{\ci\,k_{x,n}\,x}\,\frac{\cosh\{k_{n}\,(z+H)\}}{\cosh(k_{n}\,H)} 
\quad (x>\ell). 
\end{equation}
Thus ${T}^{\text{(fl)}}\equiv\vert t_{0}^{\text{(fl)}} \vert^{2}$ represents the proportion of the incident wave energy transmitted by the floe, and if the energy reflected by each floe is neglected,  the wave energy transmitted by $N$ identical floes is simply $({T}^{\text{(fl)}})^{N}$. 
The is known as the single-scattering approximation, and results in the exponential attenuation rate
\begin{equation}\label{eq:alpha_single_simple}
	\alpha = -c\,\log{({T}^{\text{(fl)}})}\,/\,\ell \approx -c\,{R}^{\text{(fl)}}\,/\,\ell,
\end{equation}
where $c$ is the concentration of floes and ${R}^{\text{(fl)}}\equiv\vert r_{0}^{\text{(fl)}} \vert^{2}=1-{T}^{\text{(fl)}}$ (by energy conservation) is the proportion of wave energy reflected by an individual floe \citep{wadhams1988attenuation}.

In general, the reflected energy increases with increasing frequency, so that the attenuation rate increases with frequency, consistent with observations.
However, the reflection coefficient experiences sharp dips at certain resonant frequencies \citep{meylan1994response}, which cause corresponding dips in the attenuation rate that do not correspond to observations.
This feature of the model can be alleviated by extending to a distribution of floe lengths and thicknesses, which 
	is incorporated in the expression for the attenuation rate (\ref{eq:alpha_single_simple}) in a straightforward manner \citep{wadhams1975airborne}.
Moreover, the expression for the predicted attenuation rate has been extended to include double scattering, which reduces the attenuation rate given by the single-scattering approximation by a factor $\approx{}2\,/\,3$ \citep{wadhams1988attenuation}.
These approximations, coupled with approximations of reflection by an individual floe, ${R}^{\text{(fl)}}$, were compared with many of the early observations by the Scott Polar Research Institute, and were found, in general, to give reasonable agreement for mid-range wave periods, where the wavelengths are comparable to the floe sizes, which is the regime in which wave scattering is expected to dominate attenuation.
For short periods the model predictions generally overpredict the observed attenuation rates and for long periods they underpredict the observations \citep{wadhams1975airborne,wadhams1978wave,wadhams1988attenuation}. 

The full solution to the problem includes all orders of multiple scattering (reflections, re-reflections, re-re-reflections, etc.). 
The resulting wave field depends on the particular realisation of the ice cover, and, hence, so does the attenuation rate.
For example, suppose the floes are identical and distributed according to some average concentration.
In the case that the floes are equally spaced (a delta-function distribution), the waves that penetrate into the marginal ice zone (beyond $x>0$) switch between frequency bands in which they propagate without attenuation and attenuate exponentially, which is the so-called passband/stopgap phenomenon known from other branches of wave science, but is not representative of waves observed in the marginal ice zone.
If the locations of the floes are randomised strongly enough (assuming the concentration allows this freedom), then waves attenuate for all frequencies, due to the Anderson localisation phenomenon \citep{bennetts2012calculation}.
For a given probability distribution, a single attenuation rate is derived by averaging over a large ensemble of solutions for randomly generated realisations of the ice cover \citep{kohout2008elastic,bennetts2012calculation}, where the averaging is with respect to wave energy rather than displacements, to avoid spurious additional attenuation due to phase cancellations.
If the randomisation is such that the waves transmitted by each floe are equally likely to have any phase, then the average attenuation rate over the ensemble can be found analytically to be $\alpha= -c\,\log({{T^{\text{(fl)}}}})\,/\,\ell$ \citep{bennetts2012calculation}, i.e., the attenuation rate is identical to the single-scattering approximation (\ref{eq:alpha_single_simple}). 
If the floes are long enough that interactions can be neglected between evanescent and decaying oscillatory waves generated at either floe end, and the floes are randomised, such that the flexural-gravity waves are equally likely to have any phase, then the expression for the attenuation rate can be reduced to
\begin{equation}\label{eq:longfloeapprox}
	\alpha = -2\,c\,\log{(1-{R}^{\text{(la)}}})\,/\,\ell,	
\end{equation}
where ${R}^{\text{(la)}}\equiv \vert {r}^{\text{(la)}}\vert^{2}$ is the proportion of incident wave energy reflected in the landfast ice problem (\S\,\ref{sec:theory:landfast}) \citep{bennetts2012calculation}.
Attenuation rate predictions from the full solution and approximation (\ref{eq:longfloeapprox}) have been compared with early observations, and, similar to the single- and double-scattering approximations, generally found to give reasonable agreement in the scattering regime but to overpredict and underpredict attenuation rates for short and long wave periods, respectively \citep{kohout2008elastic,bennetts2012model}.

Three-dimensional versions of the model have been developed, in which floes scatter waves in all directions across the ocean surface, in contrast to the two-dimensional model, which is restricted to backscattering only  \citep{bennetts2009wave,peter2010water,bennetts2010three,montiel2016attenuation}.
Studies using three-dimensional models predominantly use circular floes of uniform thickness for numerical efficiency \citep{pete2004wave}, although, in principle, floes of arbitrary shape or non-uniform thickness could be studied \citep{meylan2002wave,bennetts2010wave}.
In one class of three-dimensional model, the floes are grouped into infinite periodic rows (lines of identical and equally spaced floes), where each row shares the same periodicity \citep{bennetts2009wave,peter2010water,bennetts2010three}.
The periodicity restricts the propagating component of the wave fields scattered by each row to a small set of plane waves \citep{bennetts2010linear}, so that interactions between rows can be calculated efficiently \citep{bennetts2009wave,peter2010water}. 
Attenuation rates predicted by this model were found to give good agreement with 1979 Bering Sea observations  for wave periods from $\approx{}6$--9\,s and reasonable agreement with 1979 Greenland Sea observations  from 8--14\,s \citep{bennetts2010three}. 
Another class of three-dimensional model uses finite arrays, with no constraints on the ice floe arrangement.
Innovative computational methods are required to simulate a large enough number of floes to represent a marginal ice zone, although boundary effects still plague analysis of the outputs  \citep{montiel2016attenuation}.
Model predictions of attenuation rates were found to underpredict observations in the  Greenland Sea during the 1980s across the 5--15\,s wave period range but to predict observed directional spreading of the wave field up to a 10\,s period \citep{squire2016evolution}.

The Boltzmann equation has been used as an alternative theory to extend from the single-floe model to a model of the wave attenuation through marginal ice zone  \citep{masson1989spectral,meylan1997toward,meylan2006linear,meylan2018three}. 
The time-harmonic version of the resulting equation is of the form 
\begin{equation}\label{eq:boltzmann}
	(\cos\theta,\sin\theta)\cdot\,\nabla_{\perp}S = 
	-q\,S + \frac{c}{A_{\text{f}}}\,\int_{-\pi}^{\pi}K(\theta,\vartheta)\,S(x,y,\vartheta)\wrt\vartheta
	\mathwhere
	q = \frac{c}{A_{\text{f}}}\,\int_{-\pi}^{\pi}K(0,\vartheta)\wrt\vartheta,
\end{equation}
$A_{\text{f}}$ is the floe area, $S$ is the wave energy density at location $(x,y)$ in direction $\theta$, and $K$ is the scattering kernel derived from the single-floe model \citep{meylan1997toward,bennetts2015water}. 
It is referred to as a ``phase-averaged'' model, in contrast to ``phase-resolving'' multiple-scattering models. 
Ensemble averaging with respect to configurations is implicit in the phase averaged property of the Boltzmann model. 
Moreover, the form of (\ref{eq:boltzmann}) fits naturally into the wave energy transport equations used in most numerical ocean wave models.
(It has also been suggested that it can be approximated by an even simpler diffusion equation \citep{zhao2016diffusion}.)
However, over long distances, the Bolzmann equations predicts a steady wave field  of finite energy \citep{meylan1997toward}, which contrasts with exponential attenuation over distance predicted by phase-resolving models.

When incident wavelengths are much greater than the floe sizes, the waves ``see'' the floes as a homogenised layer on the ocean surface, rather than a collection of individual floes, and, thus, the theoretical model becomes deterministic.
In this regime, wave attenuation is dominated by dissipation of energy during wave--floe interactions, although the dominant dissipative mechanism(s) are debated.
The attenuation rate for the homogenised medium can be calculated from the classical problem (Fig.~\ref{fig:schematic-miz}) with some form of dissipation included (e.g., using Eq.~\ref{eq:dr_RP}), in the limit that the ratio of the floe size to the wavelength tends to zero \citep{pitt2024transitions}. 
The attenuation rate tends to increase as the ratio decreases, although the effect of the ice edge on the incident waves decrease, which is also indicated by physical models \citep{dolatshah2018hydroelastic,passerotti2022interactions} and provides an explanation of the observations of increased wave activity after a breakup events.

It is more common to postulate the form of the homogenised medium with one or more free parameters (usually associated to the rate of dissipation) \citep{shen2022wave}. 
This approach is likely to capture waves-in-ice physics that would not in appear in the small-floe limit of the classical model.
Calculation of the attenuation rate reduces to solving a dispersion relation to find a ``dominant'' wavenumber, 
which is usually the propagating wavenumber that has been perturbed into the complex plane by the dissipation \citep{meylan2018dispersion}.
The attenuation rate of wave energy over distance is $\alpha=-2\,\imag\{\kappa_{0}\}$, where the free parameters are usually tuned such that the attenuation rate gives a best fit to observations.

Seminal models that treat the ice layer as a viscous fluid were developed for the grease and brash ice that can occupy the outskirts of the marginal ice zone \citep{weber1987wave,keller1998gravity}.
One theory considers an asymptotically thin ice layer floating on a slightly viscous ocean (i.e., the water is no longer governed by potential-flow theory), where the viscosity in the ice layer is so large that it imposes a no-slip condition at the water surface, which creates a viscous boundary layer \citep{weber1987wave,shen2022wave}.
The resulting attenuation rate is such that
\begin{equation}
	\alpha \propto \sqrt{\nu_{\text{wtr}}}\,f^{5/2},
\end{equation}
where $\nu_{\text{wtr}}$ is the kinematic water viscosity \citep{weber1987wave}.
For appropriately selected $\nu_{\text{wtr}}$-values from 0.01--0.2\,m$^{2}$\,s$^{-1}$, the theoretical predictions were shown to give reasonable agreement with observations by the Scott Polar Research Institute in the Arctic marginal ice zone, in far more general marginal ice zone conditions than the brash/grease ice the model was designed to represent \citep{weber1987wave}.
An alternative viscous boundary layer theory for wave attenuation considers a thin elastic plate (possibly with compression), as a model of a marginal ice zone consisting of compacted ice floes, floating on viscous water \citep{liu1988wave}.
The ice cover is assumed to impose a no-slip condition on the water surface, and the resulting attenuation rate is
\begin{equation}
	\alpha \propto \frac{\kappa_{0}\,\sqrt{\nu_{\text{wtr}}\,f}}{c_{\text{g}}\,(1+\kappa_{0}\,m)}
	\mathwhere
	c_{\text{g}} = 2\,\pi\,\frac{\mathrm{d}\,f}{\mathrm{d}\kappa_{0}}
	\mathXL{is the group velocity.}
\end{equation}
The attenuation rate was found to give reasonable agreement with Arctic marginal ice zone observations for  chosen $\nu_{\text{wtr}}$-values that span a range of four orders of magnitude \citep{liu1991wave}.

Another theory for grease ice as a viscous fluid considers an ice layer to be of finite thickness and finite viscosity, floating on an inviscid ocean, so that wave attenuation occurs in the ice layer only \citep{keller1998gravity}. 
It predicts an attenuation rate such that, for long waves,
\begin{equation}
	\alpha\propto{}\nu_{\text{ice}}\,f^{5}	
\end{equation}
where $\nu_{\text{ice}}$ is the kinematic viscosity of the ice layer. 
An elastic response of the ice layer was incorporated into the theory \citep{wang2010gravity}, which connects it with the landfast ice models, although the nature of the elastic response of the ice cover for small floes in the marginal ice zone must be reinterpreted (in an unspecified manner).
The finite thickness viscoelastic model supports multiple types of wave modes that can swap dominance as parameters are varied, which makes identifying the dominant mode challenging  \citep{wang2010gravity,zhao2017nature}. 
However, for a wide parameter range, the dominant mode can be approximated by a thin plate model, which results in a dispersion relation of the form (\ref{eq:dispersion_relations}b) with a complex $F$, such that the imaginary component is proportional to frequency \citep{mosig2015comparison}.
The model was shown to give an attenuation rate within the uncertainty bounds of the observations during SIPEX~II, although using an elastic modulus several orders of magnitude greater than measured in consolidated sea ice, which has the effect of making the wavelengths in the marginal ice zone much greater than in the open ocean.
In contrast, the attenuation rate predicted by the model with damping proportional to the ice displacement velocity (\ref{eq:dr_RP}) gives comparable agreement with the SIPEX~II observations, but using an elastic modulus a couple of orders of magnitude less than that of consolidated sea ice, so that wavelengths are similar to their open-water counterparts \citep{mosig2015comparison}.

The viscous fluid ice layer theory was extended to model a mixture of grease and pancake ice \citep{de2017effect}, which is characteristic of the winter Antarctic marginal ice zone \citep{alberello2019brief} and regions of the contemporary Arctic marginal ice zone \citep{cheng2017calibrating}.
The pancakes are modelled as small rigid disks that apply no-slip conditions at the surface of the viscous fluid relative to their motion, and, thus, modify the stress at the surface of the ice layer  \citep{de2017effect}.
In conditions where the pancakes are close enough to collide, the theory is adapted to treat interacting pancakes as being locked together during the compression phase of the interaction.
The resulting ``close-packing'' theory gives an attenuation rate, such that \citep{de2017effect,shen2022wave}
\begin{equation}
\alpha\propto{} \frac{f^{5}}{\nu_{\text{ice}}},
\end{equation}
where the appearance of the viscosity parameter on the denominator contrasts with its appearance on the numerator in the theory without pancakes \citep{keller1998gravity}.
The ice-layer thickness and viscosity were estimated for the theories with and without pancakes by comparing to attenuation rate observations in grease--pancake conditions, and it was found that the viscosity parameter for the best-fits varied less for the theory with pancakes \citep{de2018ocean}.

Wave attenuation theories that go beyond the exponential attenuation paradigm have been proposed \citep{wadhams1973attenuation,shen1998wave,kohout2011wave,squire2018fresh}.
The theories proposed cover distinct mechanisms for attenuation, but they share a governing equation for the wave amplitude, $A(x)$, of the form 
\begin{equation}
\deriv{A}{x} = -\tilde{\alpha}\,A^{n},
\end{equation}
which gives exponential attenuation only in the case that $n=1$.
This class of model includes one of the earliest wave attenuation theories \citep{wadhams1973attenuation}, in which attenuation results from creep (inelastic bending) of the sea ice cover in response to waves, with the ice cover modelled as the standard floating elastic plate.
An exponent $n=3$ was chosen based on its use for ice shelves and it giving reasonable agreement with wave attenuation observations available at that time \citep{wadhams1973attenuation}.
The creep theory for wave attenuation has been rediscovered and adapted with an ad-hoc factor that reduces the attenuation when wavelengths become much greater than floe sizes \citep{boutin2018floe}, to give a model that somewhat replicates the change in wave attenuation inferred from observations before and after ice breakup events \citep{boutin2018floe,ardhuin2020ice}. 

\subsection{Waves in ice shelves}

The standard theoretical model for ocean waves propagating into and through landfast ice (\S\,\ref{sec:theory:landfast}; Fig.~\ref{fig:schematic-landfast}) has been used directly for ice shelves, although with representative geometrical parameters, i.e., thicker ice and shallower water \citep{fox1991coupling}.   
Results from the standard Kirchoff thin-plate model for floating ice were compared with results from Timoshenko--Mindlin ``thick-plate models'', finding that the additional terms in the thick-plate model, such as rotational inertia, have negligible influence in the relevant parameter ranges  \citep{fox1991coupling,balmforth1999ocean}.
Thus, there has been little subsequent interest in using thick-plate models.
The model has been extended to include the change in water depth between the open ocean and sub-shelf water cavity due to the Archimedean draught of the ice shelf, and it was used to show the draught has a major influence on model predictions over relevant wave periods, ranging from swell to tsunamis \citep{kalyanaraman2019shallow}.

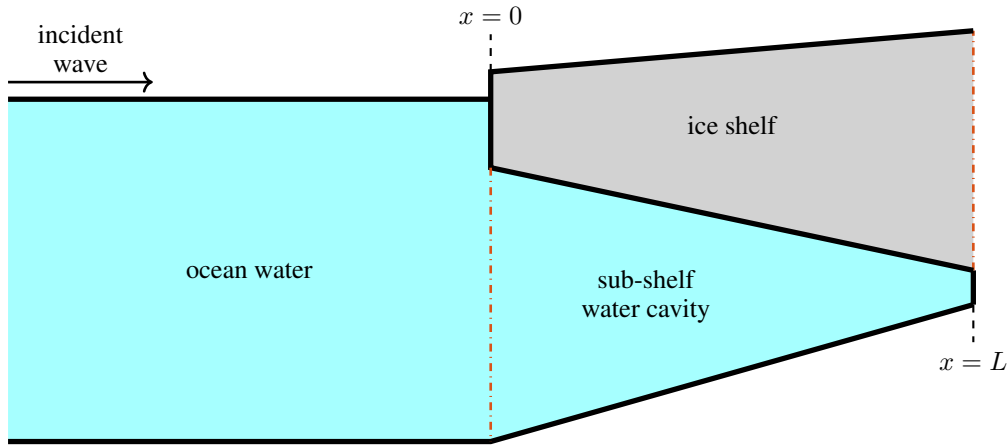
\begin{figure}[h!]
        \centering
        \setlength{\figurewidth}{0.85\textwidth} 
        \setlength{\figureheight}{0.4\textwidth} 
%
\definecolor{mycolor1}{rgb}{0.00000,1.00000,1.00000}%

\def\labelsA{1} 
\def\labelsB{0} 
\def\labelsC{0} 
\def\labelsD{1} 

\def\L{3000} 
\def\H{200}  
\def\d{100}   
\def\f{40}   

\begin{tikzpicture}

\begin{axis}[%
width=0.951\figurewidth,
height=\figureheight,
at={(0\figurewidth,0\figureheight)},
scale only axis,
xmin=-\L,
xmax=1.08*\L,
xtick={},
xlabel style={font=\color{white!15!black}},
xticklabels={},
xlabel={$x$},
ymin=-210,
ymax=80,
ytick={-200,0},
yticklabels={$-h_{0}$,0},
ylabel style={font=\color{white!15!black}},
ylabel={$z$},
axis background/.style={fill=white},
hide axis
]

\coordinate (origin) at (axis cs:0,0);
\coordinate (e1) at (axis cs:1,0);
\coordinate (e2) at (axis cs:0,1);

\coordinate (A) at (axis cs:-\L,-\H);
\coordinate (B) at (axis cs:0,-\H);
\coordinate (C) at (axis cs:\L,-0.6*\H);
\coordinate (D) at (axis cs:\L,-\d);
\coordinate (E) at (axis cs:0,-0.4*\d);
\coordinate (F) at (axis cs:-\L,0);
\coordinate (G) at (axis cs:0,0.4*\f);
\coordinate (H) at (axis cs:\L,\f);
\coordinate (I) at (axis cs:-\L,-\d);

\fill[mycolor1!35, domain=0:4000, variable=\x, samples=200] (A) -- (B) -- (C) -- (D) -- (E) -- (origin) -- (F) --cycle;

\fill[gray!35, line width=2.0pt, domain=0:4000, variable=\x, samples=200] (D) -- (E) -- (G) -- (H) -- cycle;

\draw[black, line width=2.0pt] (A) -- (B) -- (C); 
\draw[black, line width=2.0pt] (C) -- (D);
\draw[dashdotted,matlabred, line width=1.0pt] (D) -- (H);
\draw[black, line width=2.0pt] (D) -- (E) -- (G) -- (H);  
\draw[black, line width=2.0pt, domain=0:4000, variable=\x, samples=200] (F) -- (origin);
\draw[dashdotted,matlabred,line width=1.0pt] (E) -- (B);

\draw[->, black, line width=1.0pt] (axis cs: -\L,0.25*\f)  -- node[above,align=center]{incident \\ wave} (axis cs: -0.7*\L,0.25*\f);

\if\labelsA1
\node[align=center,xshift=0,yshift=-10] at ($(E)!0.5!(H)$) {ice shelf};
\node[align=center,xshift=150,yshift=-10] at ($(B)!0.5!(F)$) {sub-shelf \\ water cavity}; 
\node[align=center] at ( $ (B)!0.5!(F) $ ) {ocean water}; 
\fi
\if\labelsB1
\draw[latex-latex,black,thick] ($(B)!0.9!(A)$) -- node[right] {$H$} ($(origin)!0.9!(F)$);
\draw[latex-latex,black,thick] ($(E)!0.025!(I)$) -- node[left] {$d$} ($(origin)!0.025!(F)$);
\draw[latex-latex,black,thick] ($(E)!0.9!(D)$) -- node[left] {$h$} ($(G)!0.9!(H)$);
\fi
\if\labelsC1
\draw[-latex,thick,matlabred] (origin) -- ++(0:1.5cm) node[right] {$x$};
\draw[-latex,thick,matlabred] (origin) -- ++(90:1.5cm) node[above] {$z$};
\fi
\if\labelsD1
\draw[dashed,thick,black] (G) -- ++(90:0.5cm) node[above] {$x=0$};
\draw[dashed,thick,black] (C) -- ++(-90:0.5cm) node[below] {$x=L$};
\fi

\end{axis}

\end{tikzpicture}%
        \caption{Schematic (not to scale) of the {equilibrium} geometry for a theoretical model of ocean wave propagation into and through an ice shelf.}
        \label{fig:schematic-shelf}
\end{figure}

Another common approach is to restrict calculations to a finite interval occupied by the ice shelf and sub-shelf water cavity ($x\in[0,L]$ in Fig.~\ref{fig:schematic-shelf}).
Clamped conditions (zero displacement and slope of the ice shelf displacement) are usually applied at the grounding line, $x=L$, assuming that they represent the transition to the ice shelf becoming a grounded ice sheet for $x>L$, although hinged conditions have also been proposed \citep{holdsworth1981mechanism}.
The system is closed by prescribing (artificial) conditions along the water column beneath the ice shelf, where no water pressure \citep{holdsworth1978iceberg,holdsworth1981mechanism} and no flux (zero horizontal water velocity) \citep{sergienko2013normal,meylan2017calculation} have both been used.
Thus, the problem is unforced, and non-trivial solutions are normal modes that exist at a discrete set of wave periods and are only defined up to an unknown amplitude. 

The normal modes represent near-resonant responses of the problem in which the shelf/cavity region is connected to the open ocean, and is forced by incident waves  (Fig.~\ref{fig:schematic-shelf}) \citep{papathanasiou2019resonant}.
The connection allows resonant energy in the shelf/cavity region to leak into the open ocean, such that the normal modes for the decoupled problem  become ``complex resonances'', where the associated wave periods have imaginary components and large (but finite) responses occur for nearby real-valued wave periods \citep{bennetts2021complex}.
Depending on the parameters, particularly the wave period, the complex resonances can be better approximated by either the no-pressure or no-flux conditions \citep{bennetts2021complex}.
The problem can also be viewed as a modification of the landfast-ice-type model to have a finite length shelf/cavity region \citep{vinogradov1985oscillation,kalyanaraman2019shallow}.
Those that approach the problem from the normal mode perspective usually assume shallow-water theory, based on wavelengths in the shelf/cavity region being much greater than the cavity depth \citep{vinogradov1985oscillation,papathanasiou2019resonant}, whereas those who approach the problem as a modified version of the landfast ice problem tend to use potential-flow theory (finite depth water) \citep{ilyas2018time,meylan2021swell,bennetts2021complex}.
When the shelf/cavity region is connected to the open ocean, shallow-water theory is often inaccurate in the swell regime, as the open-ocean wavelengths are not necessarily long in relation to the water depth \citep{kalyanaraman2019shallow}.
In contrast, the ``single-mode approximation'' is accurate for the range of relevant wave periods, and has a similarly simple structure to the shallow-water approximation \citep{bennetts2021complex,liang2024pan}.

The earliest theoretical models identified the potential importance of spatial variations in the geometry (ice shelf thickness and underlying seabed) and three-dimensional effects \citep{holdsworth1978iceberg,holdsworth1981mechanism}.
However, both demand numerical solution methods, and were largely overlooked until more efficient computational approaches were developed.
Idealised spatial variations in two-dimensional geometries have been investigated, including the effects of the ice shelf thickening and the seabed shoaling away from the shelf front (Fig.~\ref{fig:schematic-shelf}) \citep{meylan2021swell,bennetts2021complex}, and blocking of waves over certain wave-period bands (i.e., stopgaps) by periodic distributions of crevasses or surface rolls \citep{freed2012blocking,nekrasov2023ocean}.
Geometries along transects through specific ice shelves have been incorported into models \citep{kalyanaraman2021icefem,bennetts2022modeling,liang2024pan}, and have been used to show that ice shelf flexure in response to swell is amplified by up to an order of magnitude at regions of local thinning (e.g., crevasses), whereas infragravity waves and very long period waves are amplified at regions of local cavity depth thinning \citep{bennetts2022modeling,liang2024pan}.
Analysis for three-dimensional models has been more restricted, but has included numerical computation of normal modes for circular, semi-circular and square ice shelves \citep{papathanasiou2019nonconforming}, and time-domain simulations for models of specific ice shelves \citep{sergienko2017behavior,tazhimbetov2023simulation}.

Theoretical models have also been investigated in which the ice shelf is treated as an elastic body of finite thickness, i.e., they do not assume the ice shelf is a plate \citep{sergienko2010elastic,sergienko2017behavior,kalyanaraman2020coupled,kalyanaraman2021icefem,abrahams2023ocean,bennetts2024thin}.  
Initial studies invoked other assumptions, such as no inertia (assuming very long waves) \citep{sergienko2010elastic,sergienko2017behavior}, or zero gravitational forcing \citep{kalyanaraman2020coupled,kalyanaraman2021icefem}.
A numerical solution method was used to conduct time-domain simulations with the full linear equations of elasticity in two dimensions and including gravitational forcing, and showed extensional Lamb waves are excited in addition to flexural-gravity waves \citep{abrahams2023ocean}.
Subsequently, a thin-plate theory was derived in which extensional Lamb waves are generated by coupling between the water and ice shelf at the shelf front \citep{bennetts2024thin}.
The new theory was used to show that extensional waves significantly increase ice shelf flexure in the swell regime in comparison to a theory with flexural-gravity waves alone \citep{bennetts2024thin}.

\section{Perspectives and outlooks}

Research to understand waves in the marginal ice zone is currently a major international and interdisciplinary research effort.
The focus on wave attenuation has persisted throughout the evolution of the research field, since it first came to prominence in the 1970s.
There are now far more observations of wave attenuation, but the central question of what mechanisms govern attenuation in the marginal ice zone remains elusive. 
The observations have served the important purpose of illustrating the challenging physics of wave attenuation in the marginal ice zone.
The associated question of how the attenuation rate depends on the ice cover properties also remains largely unresolved.
This is arguably a lower hanging fruit, as it seems likely that building on the currently limited observations of wave attenuation and accompanying ice properties \citep{alberello2022three} will reveal key relationships.
It may then be possible to limit the viable theories for wave attenuation, in a similar way to how  observed power-law relationships between the attenuation rate, $\alpha$, and wave period have been used \citep{meylan2018dispersion}.

Growth of the research fields on waves in landfast ice and ice shelves could follow that of waves in the marginal ice zone in the near future.
There is already evidence of the growth for waves in ice shelves, motivated by the increased loss of ice shelf mass to calving \citep{greene2022antarctic} and thinning \citep{paolo2015volume}, which leaves the ice shelves more susceptible to damaging wave-induced flexure \citep{bassis2024stability}.
There is also increased recognition that landfast ice has important impacts on the Earth system, despite only occupying a small fraction of the overall sea ice cover \citep{fraser2023antarctic}.
Thus, landfast ice breakup due to ocean waves is likely to play a major role in future studies, in a similar way that wave-induced breakup of large floes has played a leading role in studies of marginal ice zone dynamics over the past decade \citep{bennetts2022marginal,dumont2022marginal}.
Moreover, the three waves-in-ice sub-fields are becoming increasingly interconnected.
Large amplitude ocean swell are more likely to reach ice shelves now that the protective pack ice barrier is retreating \citep{teder2022sea}, and there is evidence that prolonged periods of flexure forced by swell triggered catastrophic calving events \citep{massom2018antarctic,teder2025large}.
Moreover, landfast ice connected to a shelf front provides an additional protective barrier from swell \citep{teder2025large}, as well as stabilising back-stress for the ice shelf \citep{greene2018seasonal}.
Therefore, there is a need to move towards research on waves in the coupled marginal ice zone--landfast ice--ice shelf system.

\section*{Acknowledgements}

This chapter is part of Comprehensive Cryospheric Science and Environmental Change.
Many thanks to Clare Eayrs for inviting the contribution and providing feedback on a draft.

\section*{Funding}

LGB is funded by the Australian Research Council (FT190100404, DP240100325).

%


\begin{thebibliography}{155}
\providecommand{\natexlab}[1]{#1}
\providecommand{\url}[1]{\texttt{#1}}
\expandafter\ifx\csname urlstyle\endcsname\relax
  \providecommand{\doi}[1]{doi: #1}\else
  \providecommand{\doi}{doi: \begingroup \urlstyle{rm}\Url}\fi

\bibitem[Bennetts et~al.(2024{\natexlab{a}})Bennetts, Shakespeare, Vreugdenhil,
  Foppert, Gayen, Meyer, Morrison, Padman, Phillips, Stevens,
  et~al.]{bennetts2024closing}
L.~G. Bennetts, C.~J. Shakespeare, C.~A. Vreugdenhil, A.~Foppert, B.~Gayen,
  A.~Meyer, A.~K. Morrison, L.~Padman, H.~E. Phillips, C.~L. Stevens, et~al.
\newblock Closing the loops on {Southern Ocean} dynamics: From the circumpolar
  current to ice shelves and from bottom mixing to surface waves.
\newblock \emph{Reviews of Geophysics}, 62\penalty0 (3):\penalty0
  e2022RG000781, 2024{\natexlab{a}}.

\bibitem[Bennetts et~al.(2022{\natexlab{a}})Bennetts, Bitz, Feltham, Kohout,
  and Meylan]{bennetts2022marginal}
L.~G. Bennetts, C.~M. Bitz, D.~L. Feltham, A.~L. Kohout, and M.~H. Meylan.
\newblock Marginal ice zone dynamics: future research perspectives and
  pathways.
\newblock \emph{Philosophical Transactions of the Royal Society A},
  380\penalty0 (2235):\penalty0 20210267, 2022{\natexlab{a}}.

\bibitem[Squire(2011)]{squire2011past}
V.~A. Squire.
\newblock Past, present and impendent hydroelastic challenges in the polar and
  subpolar seas.
\newblock \emph{Philosophical Transactions of the Royal Society A},
  369\penalty0 (1947):\penalty0 2813--2831, 2011.

\bibitem[Kohout et~al.(2014)Kohout, Williams, Dean, and
  Meylan]{kohout2014storm}
A.~L. Kohout, M.~J. Williams, S.~M. Dean, and M.~H. Meylan.
\newblock Storm-induced sea-ice breakup and the implications for ice extent.
\newblock \emph{Nature}, 509\penalty0 (7502):\penalty0 604--607, 2014.

\bibitem[Thomson et~al.(2018)Thomson, Ackley, Girard-Ardhuin, Ardhuin, Babanin,
  Boutin, Brozena, Cheng, Collins, Doble, et~al.]{thomson2018overview}
J.~Thomson, S.~Ackley, F.~Girard-Ardhuin, F.~Ardhuin, A.~Babanin, G.~Boutin,
  J.~Brozena, S.~Cheng, C.~Collins, M.~Doble, et~al.
\newblock Overview of the {A}rctic sea state and boundary layer physics
  program.
\newblock \emph{Journal of Geophysical Research: Oceans}, 123\penalty0
  (12):\penalty0 8674--8687, 2018.

\bibitem[Kohout et~al.(2020)Kohout, Smith, Roach, Williams, Montiel, and
  Williams]{kohout2020observations}
A.~L. Kohout, M.~Smith, L.~A. Roach, G.~Williams, F.~Montiel, and M.~J.
  Williams.
\newblock Observations of exponential wave attenuation in {A}ntarctic sea ice
  during the {PIPERS} campaign.
\newblock \emph{Annals of Glaciology}, 61\penalty0 (82):\penalty0 196--209,
  2020.

\bibitem[Bennetts et~al.(2022{\natexlab{b}})Bennetts, Bitz, Feltham, Kohout,
  and Meylan]{bennetts2022theory}
L.~G. Bennetts, C.~M. Bitz, D.~L. Feltham, A.~L. Kohout, and M.~H. Meylan.
\newblock Theory, modelling and observations of marginal ice zone dynamics:
  multidisciplinary perspectives and outlooks, 2022{\natexlab{b}}.

\bibitem[Squire(2022{\natexlab{a}})]{squire2022marginal}
V.~A. Squire.
\newblock Marginal ice zone dynamics.
\newblock \emph{Philosophical Transactions of the Royal Society A},
  380\penalty0 (2235):\penalty0 20210266, 2022{\natexlab{a}}.

\bibitem[Squire(2022{\natexlab{b}})]{squire2022prognosticative}
V.~A. Squire.
\newblock A prognosticative synopsis of contemporary marginal ice zone
  research.
\newblock \emph{Philosophical Transactions of the Royal Society A},
  380\penalty0 (2235):\penalty0 20220094, 2022{\natexlab{b}}.

\bibitem[Crocker and Wadhams(1989)]{crocker1989breakup}
G.~B. Crocker and P.~Wadhams.
\newblock Breakup of {A}ntarctic fast ice.
\newblock \emph{Cold Regions Science and Technology}, 17\penalty0 (1):\penalty0
  61--76, 1989.

\bibitem[Brunt et~al.(2011)Brunt, Okal, and MacAyeal]{brunt2011antarctic}
K.~M. Brunt, E.~A. Okal, and D.~R. MacAyeal.
\newblock Antarctic ice-shelf calving triggered by the {Honshu (Japan)}
  earthquake and tsunami, {M}arch 2011.
\newblock \emph{Journal of Glaciology}, 57\penalty0 (205):\penalty0 785--788,
  2011.

\bibitem[Bromirski et~al.(2010)Bromirski, Sergienko, and
  MacAyeal]{bromirski2010transoceanic}
P.~D. Bromirski, O.~V. Sergienko, and D.~R. MacAyeal.
\newblock Transoceanic infragravity waves impacting {A}ntarctic ice shelves.
\newblock \emph{Geophysical Research Letters}, 37\penalty0 (2), 2010.

\bibitem[Massom et~al.(2018)Massom, Scambos, Bennetts, Reid, Squire, and
  Stammerjohn]{massom2018antarctic}
R.~A. Massom, T.~A. Scambos, L.~G. Bennetts, P.~Reid, V.~A. Squire, and S.~E.
  Stammerjohn.
\newblock Antarctic ice shelf disintegration triggered by sea ice loss and
  ocean swell.
\newblock \emph{Nature}, 558\penalty0 (7710):\penalty0 383--389, 2018.

\bibitem[Zhao et~al.(2024)Zhao, Cheng, Fraser, Bennetts, Xiao, Liang, Li, and
  Li]{zhao2024long}
A.~Zhao, Y.~Cheng, A.~D. Fraser, L.~G. Bennetts, H.~Xiao, Q.~Liang, T.~Li, and
  R.~Li.
\newblock Long-term evolution of the {Sulzberger Ice Shelf, West Antarctica:
  I}nsights from 74-year observations and 2022 {Hunga-Tonga} volcanic
  tsunami-induced calving.
\newblock \emph{Earth and Planetary Science Letters}, 646:\penalty0 118958,
  2024.

\bibitem[Squire et~al.(1995)Squire, Dugan, Wadhams, Rottier, and
  Liu]{squire1995ocean}
V.~A. Squire, J.~P. Dugan, P.~Wadhams, P.~J. Rottier, and A.~K. Liu.
\newblock Of ocean waves and sea ice.
\newblock \emph{Annual Review of Fluid Mechanics}, 27\penalty0 (1):\penalty0
  115--168, 1995.

\bibitem[Squire(2007)]{squire2007ocean}
V.~A. Squire.
\newblock Of ocean waves and sea-ice revisited.
\newblock \emph{Cold Regions Science and Technology}, 49\penalty0 (2):\penalty0
  110--133, 2007.

\bibitem[Squire(2020)]{squire2020ocean}
V.~A. Squire.
\newblock Ocean wave interactions with sea ice: {A} reappraisal.
\newblock \emph{Annual Review of Fluid Mechanics}, 52:\penalty0 37--60, 2020.

\bibitem[Shen(2019)]{shen2019modelling}
H.~H. Shen.
\newblock Modelling ocean waves in ice-covered seas.
\newblock \emph{Applied Ocean Research}, 83:\penalty0 30--36, 2019.

\bibitem[Waseda et~al.(2022)Waseda, Alberello, Nose, Toyota, Kodaira, and
  Fujiwara]{waseda2022observation}
T.~Waseda, A.~Alberello, T.~Nose, T.~Toyota, T.~Kodaira, and Y.~Fujiwara.
\newblock Observation of anomalous spectral downshifting of waves in the
  {Okhotsk Sea} marginal ice zone.
\newblock \emph{Philosophical Transactions of the Royal Society A},
  380\penalty0 (2235):\penalty0 20210256, 2022.

\bibitem[Toffoli et~al.(2022)Toffoli, Pitt, Alberello, and
  Bennetts]{toffoli2022modelling}
A.~Toffoli, J.~P. Pitt, A.~Alberello, and L.~G. Bennetts.
\newblock Modelling attenuation of irregular wave fields by artificial ice
  floes in the laboratory.
\newblock \emph{Philosophical Transactions of the Royal Society A},
  380\penalty0 (2235):\penalty0 20210255, 2022.

\bibitem[Perrie et~al.(2022)Perrie, Meylan, Toulany, and
  Casey]{perrie2022modelling}
W.~Perrie, M.~H. Meylan, B.~Toulany, and M.~P. Casey.
\newblock Modelling wave--ice interactions in three dimensions in the marginal
  ice zone.
\newblock \emph{Philosophical Transactions of the Royal Society A},
  380\penalty0 (2235):\penalty0 20210263, 2022.

\bibitem[Shen(2022)]{shen2022wave}
H.~H. Shen.
\newblock Wave-in-ice: theoretical bases and field observations.
\newblock \emph{Philosophical Transactions of the Royal Society A},
  380\penalty0 (2235):\penalty0 20210254, 2022.

\bibitem[Thomson(2022)]{thomson2022wave}
J.~Thomson.
\newblock Wave propagation in the marginal ice zone: connections and feedback
  mechanisms within the air--ice--ocean system.
\newblock \emph{Philosophical Transactions of the Royal Society A},
  380\penalty0 (2235):\penalty0 20210251, 2022.

\bibitem[Golden et~al.(2020)Golden, Bennetts, Cherkaev, Eisenman, Feltham,
  Horvat, Hunke, Jones, Perovich, Ponte-Castaneda, et~al.]{golden2020modeling}
K.~M. Golden, L.~G. Bennetts, E.~Cherkaev, I.~Eisenman, D.~Feltham, C.~Horvat,
  E.~Hunke, C.~Jones, D.~K. Perovich, P.~Ponte-Castaneda, et~al.
\newblock Modeling sea ice.
\newblock \emph{Notices of the American Mathematical Society}, 67\penalty0
  (10):\penalty0 1535--1555, 2020.

\bibitem[Fraser et~al.(2023)Fraser, Wongpan, Langhorne, Klekociuk, Kusahara,
  Lannuzel, Massom, Meiners, Swadling, Atwater, et~al.]{fraser2023antarctic}
A.~D. Fraser, P.~Wongpan, P.~J. Langhorne, A.~Klekociuk, K.~Kusahara,
  D.~Lannuzel, R.~A. Massom, K.~M. Meiners, K.~Swadling, D.~P. Atwater, et~al.
\newblock Antarctic landfast sea ice: {A} review of its physics,
  biogeochemistry and ecology.
\newblock \emph{Reviews of Geophysics}, 61\penalty0 (2):\penalty0
  e2022RG000770, 2023.

\bibitem[Squire and Allan(1977)]{squire1977propagation}
V.~A. Squire and A.~Allan.
\newblock \emph{Propagation of flexural gravity waves in sea ice}.
\newblock Centre for Cold Ocean Resources Engineering, Memorial University of
  Newfoundland, 1977.

\bibitem[Squire(1984{\natexlab{a}})]{squire1984theoretical}
V.~A. Squire.
\newblock A theoretical, laboratory, and field study of ice-coupled waves.
\newblock \emph{Journal of Geophysical Research: Oceans}, 89\penalty0
  (C5):\penalty0 8069--8079, 1984{\natexlab{a}}.

\bibitem[Sutherland and Rabault(2016)]{sutherland2016observations}
G.~Sutherland and J.~Rabault.
\newblock Observations of wave dispersion and attenuation in landfast ice.
\newblock \emph{Journal of Geophysical Research: Oceans}, 121\penalty0
  (3):\penalty0 1984--1997, 2016.

\bibitem[Voermans et~al.(2021)Voermans, Liu, Marchenko, Rabault, Filchuk,
  Ryzhov, Heil, Waseda, Nose, Kodaira, et~al.]{voermans2021wave}
J.~J. Voermans, Q.~Liu, A.~Marchenko, J.~Rabault, K.~Filchuk, I.~Ryzhov,
  P.~Heil, T.~Waseda, T.~Nose, T.~Kodaira, et~al.
\newblock Wave dispersion and dissipation in landfast ice: comparison of
  observations against models.
\newblock \emph{The Cryosphere}, 15\penalty0 (12):\penalty0 5557--5575, 2021.

\bibitem[Robin(1963)]{robin1963wave}
G.~d.~Q. Robin.
\newblock Wave propagation through fields of pack ice.
\newblock \emph{Philosophical Transactions of the Royal Society A},
  255\penalty0 (1057):\penalty0 313--339, 1963.

\bibitem[Wadhams(1975)]{wadhams1975airborne}
P.~Wadhams.
\newblock Airborne laser profiling of swell in an open ice field.
\newblock \emph{Journal of Geophysical Research}, 80\penalty0 (33):\penalty0
  4520--4528, 1975.

\bibitem[Wadhams(1978)]{wadhams1978wave}
P.~Wadhams.
\newblock Wave decay in the marginal ice zone measured from a submarine.
\newblock \emph{Deep Sea Research}, 25\penalty0 (1):\penalty0 23--40, 1978.

\bibitem[Squire and Moore(1980)]{squire1980direct}
V.~A. Squire and S.~C. Moore.
\newblock Direct measurement of the attenuation of ocean waves by pack ice.
\newblock \emph{Nature}, 283\penalty0 (5745):\penalty0 365--368, 1980.

\bibitem[Wadhams(1985)]{wadhams1985marginal}
P.~Wadhams.
\newblock The marginal ice zone experiment {(MIZEX) 1984: Scott Polar Research
  Institute} participation.
\newblock \emph{Polar Record}, 22\penalty0 (140):\penalty0 505--510, 1985.

\bibitem[Wadhams et~al.(1986)Wadhams, Squire, Ewing, and
  Pascal]{wadhams1986effect}
P.~Wadhams, V.~A. Squire, J.~Ewing, and R.~Pascal.
\newblock The effect of the marginal ice zone on the directional wave spectrum
  of the ocean.
\newblock \emph{Journal of Physical Oceanography}, 16\penalty0 (2):\penalty0
  358--376, 1986.

\bibitem[Wadhams et~al.(1988)Wadhams, Squire, Goodman, Cowan, and
  Moore]{wadhams1988attenuation}
P.~Wadhams, V.~A. Squire, D.~J. Goodman, A.~M. Cowan, and S.~C. Moore.
\newblock The attenuation rates of ocean waves in the marginal ice zone.
\newblock \emph{Journal of Geophysical Research: Oceans}, 93\penalty0
  (C6):\penalty0 6799--6818, 1988.

\bibitem[Wadhams and Holt(1991)]{wadhams1991waves}
P.~Wadhams and B.~Holt.
\newblock Waves in frazil and pancake ice and their detection in {Seasat}
  synthetic aperture radar imagery.
\newblock \emph{Journal of Geophysical Research: Oceans}, 96\penalty0
  (C5):\penalty0 8835--8852, 1991.

\bibitem[Liu et~al.(1991{\natexlab{a}})Liu, Holt, and Vachon]{liu1991wave}
A.~K. Liu, B.~Holt, and P.~W. Vachon.
\newblock Wave propagation in the marginal ice zone: {M}odel predictions and
  comparisons with buoy and synthetic aperture radar data.
\newblock \emph{Journal of Geophysical Research: Oceans}, 96\penalty0
  (C3):\penalty0 4605--4621, 1991{\natexlab{a}}.

\bibitem[Liu et~al.(1991{\natexlab{b}})Liu, Vachon, and
  Peng]{liu1991observation}
A.~K. Liu, P.~W. Vachon, and C.~Y. Peng.
\newblock Observation of wave refraction at an ice edge by synthetic aperture
  radar.
\newblock \emph{Journal of Geophysical Research: Oceans}, 96\penalty0
  (C3):\penalty0 4803--4808, 1991{\natexlab{b}}.

\bibitem[Larouche and Cariou(1992)]{larouche1992directional}
P.~Larouche and C.~Cariou.
\newblock Directional wave spectra estimation in a marginal ice zone using
  linear prediction.
\newblock \emph{Journal of Physical Oceanography}, 22\penalty0 (2):\penalty0
  196--206, 1992.

\bibitem[Wadhams et~al.(2002)Wadhams, Parmiggiani, and
  De~Carolis]{wadhams2002use}
P.~Wadhams, F.~Parmiggiani, and G.~De~Carolis.
\newblock The use of {SAR} to measure ocean wave dispersion in frazil--pancake
  icefields.
\newblock \emph{Journal of Physical Oceanography}, 32\penalty0 (6):\penalty0
  1721--1746, 2002.

\bibitem[Wadhams et~al.(2004)Wadhams, Parmiggiani, De~Carolis, Desiderio, and
  Doble]{wadhams2004sar}
P.~Wadhams, F.~Parmiggiani, G.~De~Carolis, D.~Desiderio, and M.~J. Doble.
\newblock {SAR} imaging of wave dispersion in {A}ntarctic pancake ice and its
  use in measuring ice thickness.
\newblock \emph{Geophysical Research Letters}, 31\penalty0 (15), 2004.

\bibitem[Doble and Bidlot(2013)]{doble2013wave}
M.~J. Doble and J.-R. Bidlot.
\newblock Wave buoy measurements at the antarctic sea ice edge compared with an
  enhanced {ECMWF WAM: P}rogress towards global waves-in-ice modelling.
\newblock \emph{Ocean Modelling}, 70:\penalty0 166--173, 2013.

\bibitem[Doble et~al.(2015)Doble, De~Carolis, Meylan, Bidlot, and
  Wadhams]{doble2015relating}
M.~J. Doble, G.~De~Carolis, M.~H. Meylan, J.-R. Bidlot, and P.~Wadhams.
\newblock Relating wave attenuation to pancake ice thickness, using field
  measurements and model results.
\newblock \emph{Geophysical Research Letters}, 42\penalty0 (11):\penalty0
  4473--4481, 2015.

\bibitem[Meylan et~al.(2014)Meylan, Bennetts, and Kohout]{meylan2014situ}
M.~H. Meylan, L.~G. Bennetts, and A.~L. Kohout.
\newblock In situ measurements and analysis of ocean waves in the {A}ntarctic
  marginal ice zone.
\newblock \emph{Geophysical Research Letters}, 41\penalty0 (14):\penalty0
  5046--5051, 2014.

\bibitem[Cheng et~al.(2017)Cheng, Rogers, Thomson, Smith, Doble, Wadhams,
  Kohout, Lund, Persson, Collins~III, et~al.]{cheng2017calibrating}
S.~Cheng, W.~E. Rogers, J.~Thomson, M.~Smith, M.~J. Doble, P.~Wadhams, A.~L.
  Kohout, B.~Lund, O.~P. Persson, C.~O. Collins~III, et~al.
\newblock Calibrating a viscoelastic sea ice model for wave propagation in the
  {A}rctic fall marginal ice zone.
\newblock \emph{Journal of Geophysical Research: Oceans}, 122\penalty0
  (11):\penalty0 8770--8793, 2017.

\bibitem[Collins et~al.(2018)Collins, Doble, Lund, and
  Smith]{collins2018observations}
C.~Collins, M.~Doble, B.~Lund, and M.~Smith.
\newblock Observations of surface wave dispersion in the marginal ice zone.
\newblock \emph{Journal of Geophysical Research: Oceans}, 123\penalty0
  (5):\penalty0 3336--3354, 2018.

\bibitem[Montiel et~al.(2018)Montiel, Squire, Doble, Thomson, and
  Wadhams]{montiel2018attenuation}
F.~Montiel, V.~Squire, M.~Doble, J.~Thomson, and P.~Wadhams.
\newblock Attenuation and directional spreading of ocean waves during a storm
  event in the autumn {Beaufort Sea} marginal ice zone.
\newblock \emph{Journal of Geophysical Research: Oceans}, 123\penalty0
  (8):\penalty0 5912--5932, 2018.

\bibitem[Rogers et~al.(2021)Rogers, Meylan, and Kohout]{rogers2021estimates}
W.~E. Rogers, M.~H. Meylan, and A.~L. Kohout.
\newblock Estimates of spectral wave attenuation in {A}ntarctic sea ice, using
  model/data inversion.
\newblock \emph{Cold Regions Science and Technology}, 182:\penalty0 103198,
  2021.

\bibitem[Montiel et~al.(2022)Montiel, Kohout, and Roach]{montiel2022physical}
F.~Montiel, A.~L. Kohout, and L.~A. Roach.
\newblock Physical drivers of ocean wave attenuation in the marginal ice zone.
\newblock \emph{Journal of Physical Oceanography}, 52\penalty0 (5):\penalty0
  889--906, 2022.

\bibitem[Vichi et~al.(2019)Vichi, Eayrs, Alberello, Bekker, Bennetts, Holland,
  de~Jong, Joubert, MacHutchon, Messori, et~al.]{vichi2019effects}
M.~Vichi, C.~Eayrs, A.~Alberello, A.~Bekker, L.~G. Bennetts, D.~Holland,
  E.~de~Jong, W.~Joubert, K.~MacHutchon, G.~Messori, et~al.
\newblock Effects of an explosive polar cyclone crossing the {A}ntarctic
  marginal ice zone.
\newblock \emph{Geophysical Research Letters}, 46\penalty0 (11):\penalty0
  5948--5958, 2019.

\bibitem[Alberello et~al.(2020)Alberello, Bennetts, Heil, Eayrs, Vichi,
  MacHutchon, Onorato, and Toffoli]{alberello2020drift}
A.~Alberello, L.~Bennetts, P.~Heil, C.~Eayrs, M.~Vichi, K.~MacHutchon,
  M.~Onorato, and A.~Toffoli.
\newblock Drift of pancake ice floes in the winter {Antarctic} marginal ice
  zone during polar cyclones.
\newblock \emph{Journal of Geophysical Research: Oceans}, 125\penalty0
  (3):\penalty0 e2019JC015418, 2020.

\bibitem[Nose et~al.(2024)Nose, Katsuno, Waseda, Ushio, Rabault, Kodaira, and
  Voermans]{nose2024observation}
T.~Nose, T.~Katsuno, T.~Waseda, S.~Ushio, J.~Rabault, T.~Kodaira, and
  J.~Voermans.
\newblock Observation of wave propagation over 1,000 km into {A}ntarctica
  winter pack ice.
\newblock \emph{Coastal Engineering Journal}, 66\penalty0 (1):\penalty0
  115--131, 2024.

\bibitem[Ardhuin et~al.(2020)Ardhuin, Otero, Merrifield, Grouazel, and
  Terrill]{ardhuin2020ice}
F.~Ardhuin, M.~Otero, S.~Merrifield, A.~Grouazel, and E.~Terrill.
\newblock Ice breakup controls dissipation of wind waves across {Southern
  Ocean} sea ice.
\newblock \emph{Geophysical Research Letters}, 47\penalty0 (13):\penalty0
  e2020GL087699, 2020.

\bibitem[Sutherland and Gascard(2016)]{sutherland2016airborne}
P.~Sutherland and J.-C. Gascard.
\newblock Airborne remote sensing of ocean wave directional wavenumber spectra
  in the marginal ice zone.
\newblock \emph{Geophysical Research Letters}, 43\penalty0 (10):\penalty0
  5151--5159, 2016.

\bibitem[Sutherland et~al.(2018)Sutherland, Brozena, Rogers, Doble, and
  Wadhams]{sutherland2018airborne}
P.~Sutherland, J.~Brozena, W.~E. Rogers, M.~Doble, and P.~Wadhams.
\newblock Airborne remote sensing of wave propagation in the marginal ice zone.
\newblock \emph{Journal of Geophysical Research: Oceans}, 123\penalty0
  (6):\penalty0 4132--4152, 2018.

\bibitem[Ardhuin et~al.(2015)Ardhuin, Collard, Chapron, Girard-Ardhuin,
  Guitton, Mouche, and Stopa]{ardhuin2015estimates}
F.~Ardhuin, F.~Collard, B.~Chapron, F.~Girard-Ardhuin, G.~Guitton, A.~Mouche,
  and J.~E. Stopa.
\newblock Estimates of ocean wave heights and attenuation in sea ice using the
  {SAR} wave mode on {Sentinel-1A}.
\newblock \emph{Geophysical Research Letters}, 42\penalty0 (7):\penalty0
  2317--2325, 2015.

\bibitem[Ardhuin et~al.(2017)Ardhuin, Stopa, Chapron, Collard, Smith, Thomson,
  Doble, Blomquist, Persson, Collins~III, et~al.]{ardhuin2017measuring}
F.~Ardhuin, J.~Stopa, B.~Chapron, F.~Collard, M.~Smith, J.~Thomson, M.~Doble,
  B.~Blomquist, O.~Persson, C.~O. Collins~III, et~al.
\newblock Measuring ocean waves in sea ice using {SAR} imagery: {A}
  quasi-deterministic approach evaluated with {Sentinel-1} and in situ data.
\newblock \emph{Remote sensing of Environment}, 189:\penalty0 211--222, 2017.

\bibitem[Stopa et~al.(2018{\natexlab{a}})Stopa, Ardhuin, Thomson, Smith,
  Kohout, Doble, and Wadhams]{stopa2018wave}
J.~E. Stopa, F.~Ardhuin, J.~Thomson, M.~M. Smith, A.~Kohout, M.~Doble, and
  P.~Wadhams.
\newblock Wave attenuation through an {A}rctic marginal ice zone on 12
  {O}ctober 2015: 1. {M}easurement of wave spectra and ice features from
  {Sentinel 1A}.
\newblock \emph{Journal of Geophysical Research: Oceans}, 123\penalty0
  (5):\penalty0 3619--3634, 2018{\natexlab{a}}.

\bibitem[Stopa et~al.(2018{\natexlab{b}})Stopa, Sutherland, and
  Ardhuin]{stopa2018strong}
J.~E. Stopa, P.~Sutherland, and F.~Ardhuin.
\newblock Strong and highly variable push of ocean waves on {Southern Ocean}
  sea ice.
\newblock \emph{Proceedings of the National Academy of Sciences}, 115\penalty0
  (23):\penalty0 5861--5865, 2018{\natexlab{b}}.

\bibitem[Huang and Li(2023)]{huang2023wave}
B.~Q. Huang and X.-M. Li.
\newblock Wave attenuation by sea ice in the {A}rctic marginal ice zone
  observed by spaceborne {SAR}.
\newblock \emph{Geophysical Research Letters}, 50\penalty0 (21):\penalty0
  e2023GL105059, 2023.

\bibitem[Horvat et~al.(2020)Horvat, Blanchard-Wrigglesworth, and
  Petty]{horvat2020observing}
C.~Horvat, E.~Blanchard-Wrigglesworth, and A.~Petty.
\newblock Observing waves in sea ice with {ICESat-2}.
\newblock \emph{Geophysical Research Letters}, 47\penalty0 (10):\penalty0
  e2020GL087629, 2020.

\bibitem[Brouwer et~al.(2022)Brouwer, Fraser, Murphy, Wongpan, Alberello,
  Kohout, Horvat, Wotherspoon, Massom, Cartwright,
  et~al.]{brouwer2022altimetric}
J.~Brouwer, A.~D. Fraser, D.~J. Murphy, P.~Wongpan, A.~Alberello, A.~Kohout,
  C.~Horvat, S.~Wotherspoon, R.~A. Massom, J.~Cartwright, et~al.
\newblock Altimetric observation of wave attenuation through the {A}ntarctic
  marginal ice zone using {ICESat-2}.
\newblock \emph{The Cryosphere}, 16\penalty0 (6):\penalty0 2325--2353, 2022.

\bibitem[Hell and Horvat(2024)]{hell2024method}
M.~C. Hell and C.~Horvat.
\newblock A method for constructing directional surface wave spectra from
  {ICESat-2} altimetry.
\newblock \emph{The Cryosphere}, 18\penalty0 (1):\penalty0 341--361, 2024.

\bibitem[Thiel et~al.(1960)Thiel, Crary, Haubrich, and
  Behrendt]{thiel1960gravimetric}
E.~Thiel, A.~Crary, R.~A. Haubrich, and J.~C. Behrendt.
\newblock Gravimetric determination of ocean tide, {Weddell and Ross Seas,
  Antarctica}.
\newblock \emph{Journal of Geophysical Research}, 65\penalty0 (2):\penalty0
  629--636, 1960.

\bibitem[Williams and Robinson(1981)]{williams1981flexural}
R.~Williams and E.~Robinson.
\newblock Flexural waves in the {Ross Ice Shelf}.
\newblock \emph{Journal of Geophysical Research: Oceans}, 86\penalty0
  (C7):\penalty0 6643--6648, 1981.

\bibitem[Robinson and Haskell(1992)]{robinson1992travelling}
W.~Robinson and T.~Haskell.
\newblock Travelling flexural waves in the {Erebus Glacier Tongue, McMurdo
  Sound, Antarctica}.
\newblock \emph{Cold Regions Science and Technology}, 20\penalty0 (3):\penalty0
  289--293, 1992.

\bibitem[Squire et~al.(1994)Squire, Robinson, Meylan, and
  Haskell]{squire1994observations}
V.~A. Squire, W.~H. Robinson, M.~Meylan, and T.~G. Haskell.
\newblock Observations of flexural waves on the {Erebus Ice Tongue, McMurdo
  Sound, Antarctica}, and nearby sea ice.
\newblock \emph{Journal of Glaciology}, 40\penalty0 (135):\penalty0 377--385,
  1994.

\bibitem[Robinson and Haskell(1990)]{robinson1990calving}
W.~H. Robinson and T.~G. Haskell.
\newblock Calving of {E}rebus {G}lacier {T}ongue.
\newblock \emph{Nature}, 346\penalty0 (6285):\penalty0 615--616, 1990.

\bibitem[MacAyeal et~al.(2006)MacAyeal, Okal, Aster, Bassis, Brunt, Cathles,
  Drucker, Fricker, Kim, Martin, et~al.]{macayeal2006transoceanic}
D.~R. MacAyeal, E.~A. Okal, R.~C. Aster, J.~N. Bassis, K.~M. Brunt, L.~M.
  Cathles, R.~Drucker, H.~A. Fricker, Y.-J. Kim, S.~Martin, et~al.
\newblock Transoceanic wave propagation links iceberg calving margins of
  {A}ntarctica with storms in tropics and {N}orthern {H}emisphere.
\newblock \emph{Geophysical Research Letters}, 33\penalty0 (17), 2006.

\bibitem[Cathles~IV et~al.(2009)Cathles~IV, Okal, and
  MacAyeal]{cathles2009seismic}
L.~Cathles~IV, E.~A. Okal, and D.~R. MacAyeal.
\newblock Seismic observations of sea swell on the floating {Ross Ice Shelf,
  Antarctica}.
\newblock \emph{Journal of Geophysical Research: Earth Surface}, 114\penalty0
  (F2), 2009.

\bibitem[Bromirski and Stephen(2012)]{bromirski2012response}
P.~D. Bromirski and R.~A. Stephen.
\newblock Response of the {Ross Ice Shelf, Antarctica}, to ocean gravity-wave
  forcing.
\newblock \emph{Annals of Glaciology}, 53\penalty0 (60):\penalty0 163--172,
  2012.

\bibitem[Bromirski et~al.(2015)Bromirski, Diez, Gerstoft, Stephen, Bolmer,
  Wiens, Aster, and Nyblade]{bromirski2015ross}
P.~D. Bromirski, A.~Diez, P.~Gerstoft, R.~A. Stephen, T.~Bolmer, D.~Wiens,
  R.~Aster, and A.~Nyblade.
\newblock {Ross Ice Shelf} vibrations.
\newblock \emph{Geophysical Research Letters}, 42\penalty0 (18):\penalty0
  7589--7597, 2015.

\bibitem[Bromirski et~al.(2017)Bromirski, Chen, Stephen, Gerstoft, Arcas, Diez,
  Aster, Wiens, and Nyblade]{bromirski2017tsunami}
P.~D. Bromirski, Z.~Chen, R.~A. Stephen, P.~Gerstoft, D.~Arcas, A.~Diez, R.~C.
  Aster, D.~A. Wiens, and A.~Nyblade.
\newblock Tsunami and infragravity waves impacting {A}ntarctic ice shelves.
\newblock \emph{Journal of Geophysical Research: Oceans}, 122\penalty0
  (7):\penalty0 5786--5801, 2017.

\bibitem[Chen et~al.(2018)Chen, Bromirski, Gerstoft, Stephen, Wiens, Aster, and
  Nyblade]{chen2018ocean}
Z.~Chen, P.~D. Bromirski, P.~Gerstoft, R.~A. Stephen, D.~A. Wiens, R.~C. Aster,
  and A.~A. Nyblade.
\newblock Ocean-excited plate waves in the {Ross and Pine Island Glacier} ice
  shelves.
\newblock \emph{Journal of Glaciology}, 64\penalty0 (247):\penalty0 730--744,
  2018.

\bibitem[Chen et~al.(2019)Chen, Bromirski, Gerstoft, Stephen, Lee, Yun,
  Olinger, Aster, Wiens, and Nyblade]{chen2019ross}
Z.~Chen, P.~D. Bromirski, P.~Gerstoft, R.~A. Stephen, W.~S. Lee, S.~Yun, S.~D.
  Olinger, R.~C. Aster, D.~A. Wiens, and A.~A. Nyblade.
\newblock {Ross Ice Shelf} icequakes associated with ocean gravity wave
  activity.
\newblock \emph{Geophysical Research Letters}, 46\penalty0 (15):\penalty0
  8893--8902, 2019.

\bibitem[Fox and Squire(1990)]{fox1990reflection}
C.~Fox and V.~A. Squire.
\newblock Reflection and transmission characteristics at the edge of shore fast
  sea ice.
\newblock \emph{Journal of Geophysical Research: Oceans}, 95\penalty0
  (C7):\penalty0 11629--11639, 1990.

\bibitem[Fox and Squire(1994)]{fox1994oblique}
C.~Fox and V.~A. Squire.
\newblock On the oblique reflexion and transmission of ocean waves at shore
  fast sea ice.
\newblock \emph{Philosophical Transactions of the Royal Society A},
  347\penalty0 (1682):\penalty0 185--218, 1994.

\bibitem[Chung and Fox(2001)]{chung2001calculation}
H.~Chung and C.~Fox.
\newblock Calculation of wave propagation into land-fast ice.
\newblock \emph{Annals of Glaciology}, 33:\penalty0 322--326, 2001.

\bibitem[Linton and McIver(2001)]{linton2001handbook}
C.~M. Linton and P.~McIver.
\newblock \emph{Handbook of mathematical techniques for wave/structure
  interactions}.
\newblock Chapman and Hall/CRC, 2001.

\bibitem[Williams(2006)]{williams2006reflections}
T.~D. Williams.
\newblock \emph{Reflections on ice: Scattering of Flexural Gravity}.
\newblock PhD thesis, University of Otago, 2006.

\bibitem[Bennetts(2007)]{bennetts2007wave}
L.~G. Bennetts.
\newblock \emph{Wave scattering by ice sheets of varying thickness}.
\newblock PhD thesis, University of Reading, 2007.

\bibitem[Squire(1984{\natexlab{b}})]{squire1984critical}
V.~A. Squire.
\newblock On the critical angle for ocean waves entering shore fast ice.
\newblock \emph{Cold Regions Science and Technology}, 10\penalty0 (1):\penalty0
  59--68, 1984{\natexlab{b}}.

\bibitem[Robinson and Palmer(1990)]{robinson1990modal}
N.~Robinson and S.~Palmer.
\newblock A modal analysis of a rectangular plate floating on an incompressible
  liquid.
\newblock \emph{Journal of Sound and Vibration}, 142\penalty0 (3):\penalty0
  453--460, 1990.

\bibitem[Squire and Fox(1992)]{squire1992ice}
V.~A. Squire and C.~Fox.
\newblock On ice coupled waves: a comparison of data and theory.
\newblock In \emph{Advances in ice technology: Proc. 3rd Int. Conf. on Ice
  Technology}, pages 269--280. Computational Mechanics Publications Cambridge,
  MA, 1992.

\bibitem[Squire(1993)]{squire1993breakup}
V.~A. Squire.
\newblock The breakup of shore fast sea ice.
\newblock \emph{Cold Regions Science and Technology}, 21\penalty0 (3):\penalty0
  211--218, 1993.

\bibitem[Meylan and Squire(1994)]{meylan1994response}
M.~H. Meylan and V.~A. Squire.
\newblock The response of ice floes to ocean waves.
\newblock \emph{Journal of Geophysical Research: Oceans}, 99\penalty0
  (C1):\penalty0 891--900, 1994.

\bibitem[Bennetts and Squire(2012{\natexlab{a}})]{bennetts2012calculation}
L.~G. Bennetts and V.~A. Squire.
\newblock On the calculation of an attenuation coefficient for transects of
  ice-covered ocean.
\newblock \emph{Proceedings of the Royal Society A}, 468\penalty0
  (2137):\penalty0 136--162, 2012{\natexlab{a}}.

\bibitem[Kohout and Meylan(2008)]{kohout2008elastic}
A.~L. Kohout and M.~H. Meylan.
\newblock An elastic plate model for wave attenuation and ice floe breaking in
  the marginal ice zone.
\newblock \emph{Journal of Geophysical Research: Oceans}, 113\penalty0 (C9),
  2008.

\bibitem[Bennetts and Squire(2012{\natexlab{b}})]{bennetts2012model}
L.~G. Bennetts and V.~A. Squire.
\newblock Model sensitivity analysis of scattering-induced attenuation of
  ice-coupled waves.
\newblock \emph{Ocean Modelling}, 45:\penalty0 1--13, 2012{\natexlab{b}}.

\bibitem[Bennetts and Squire(2009)]{bennetts2009wave}
L.~Bennetts and V.~Squire.
\newblock Wave scattering by multiple rows of circular ice floes.
\newblock \emph{Journal of Fluid Mechanics}, 639:\penalty0 213--238, 2009.

\bibitem[Peter and Meylan(2010)]{peter2010water}
M.~A. Peter and M.~H. Meylan.
\newblock Water-wave scattering by vast fields of bodies.
\newblock \emph{SIAM Journal on Applied Mathematics}, 70\penalty0 (5):\penalty0
  1567--1586, 2010.

\bibitem[Bennetts et~al.(2010)Bennetts, Peter, Squire, and
  Meylan]{bennetts2010three}
L.~G. Bennetts, M.~A. Peter, V.~Squire, and M.~H. Meylan.
\newblock A three-dimensional model of wave attenuation in the marginal ice
  zone.
\newblock \emph{Journal of Geophysical Research: Oceans}, 115\penalty0 (C12),
  2010.

\bibitem[Montiel et~al.(2016)Montiel, Squire, and
  Bennetts]{montiel2016attenuation}
F.~Montiel, V.~Squire, and L.~G. Bennetts.
\newblock Attenuation and directional spreading of ocean wave spectra in the
  marginal ice zone.
\newblock \emph{Journal of Fluid Mechanics}, 790:\penalty0 492--522, 2016.

\bibitem[Peter et~al.(2004)Peter, Meylan, and Chung]{pete2004wave}
M.~A. Peter, M.~H. Meylan, and H.~Chung.
\newblock Wave scattering by a circular elastic plate in water of finite depth:
  a closed form solution.
\newblock \emph{International Journal of Offshore and Polar Engineering},
  14\penalty0 (02), 2004.

\bibitem[Meylan(2002)]{meylan2002wave}
M.~H. Meylan.
\newblock Wave response of an ice floe of arbitrary geometry.
\newblock \emph{Journal of Geophysical Research: Oceans}, 107\penalty0
  (C1):\penalty0 5--1, 2002.

\bibitem[Bennetts and Williams(2010)]{bennetts2010wave}
L.~G. Bennetts and T.~D. Williams.
\newblock Wave scattering by ice floes and polynyas of arbitrary shape.
\newblock \emph{Journal of Fluid Mechanics}, 662:\penalty0 5--35, 2010.

\bibitem[Bennetts and Squire(2010)]{bennetts2010linear}
L.~G. Bennetts and V.~A. Squire.
\newblock Linear wave forcing of an array of axisymmetric ice floes.
\newblock \emph{IMA Journal of Applied Mathematics}, 75\penalty0 (1):\penalty0
  108--138, 2010.

\bibitem[Squire and Montiel(2016)]{squire2016evolution}
V.~A. Squire and F.~Montiel.
\newblock Evolution of directional wave spectra in the marginal ice zone: a new
  model tested with legacy data.
\newblock \emph{Journal of Physical Oceanography}, 46\penalty0 (10):\penalty0
  3121--3137, 2016.

\bibitem[Masson and LeBlond(1989)]{masson1989spectral}
D.~Masson and P.~LeBlond.
\newblock Spectral evolution of wind-generated surface gravity waves in a
  dispersed ice field.
\newblock \emph{Journal of Fluid Mechanics}, 202:\penalty0 43--81, 1989.

\bibitem[Meylan et~al.(1997)Meylan, Squire, and Fox]{meylan1997toward}
M.~H. Meylan, V.~A. Squire, and C.~Fox.
\newblock Toward realism in modeling ocean wave behavior in marginal ice zones.
\newblock \emph{Journal of Geophysical Research: Oceans}, 102\penalty0
  (C10):\penalty0 22981--22991, 1997.

\bibitem[Meylan and Masson(2006)]{meylan2006linear}
M.~H. Meylan and D.~Masson.
\newblock A linear {B}oltzmann equation to model wave scattering in the
  marginal ice zone.
\newblock \emph{Ocean Modelling}, 11\penalty0 (3-4):\penalty0 417--427, 2006.

\bibitem[Meylan and Bennetts(2018)]{meylan2018three}
M.~H. Meylan and L.~G. Bennetts.
\newblock Three-dimensional time-domain scattering of waves in the marginal ice
  zone.
\newblock \emph{Philosophical Transactions of the Royal Society A},
  376\penalty0 (2129):\penalty0 20170334, 2018.

\bibitem[Bennetts and Williams(2015)]{bennetts2015water}
L.~G. Bennetts and T.~D. Williams.
\newblock Water wave transmission by an array of floating discs.
\newblock \emph{Proceedings of the Royal Society A}, 471\penalty0
  (2173):\penalty0 20140698, 2015.

\bibitem[Zhao and Shen(2016)]{zhao2016diffusion}
X.~Zhao and H.~H. Shen.
\newblock A diffusion approximation for ocean wave scatterings by randomly
  distributed ice floes.
\newblock \emph{Ocean Modelling}, 107:\penalty0 21--27, 2016.

\bibitem[Pitt and Bennetts(2024)]{pitt2024transitions}
J.~P. Pitt and L.~G. Bennetts.
\newblock On transitions in water wave propagation through consolidated to
  broken sea ice covers.
\newblock \emph{Proceedings of the Royal Society A}, 480\penalty0
  (2297):\penalty0 20230862, 2024.

\bibitem[Dolatshah et~al.(2018)Dolatshah, Nelli, Bennetts, Alberello, Meylan,
  Monty, and Toffoli]{dolatshah2018hydroelastic}
A.~Dolatshah, F.~Nelli, L.~G. Bennetts, A.~Alberello, M.~H. Meylan, J.~P.
  Monty, and A.~Toffoli.
\newblock Hydroelastic interactions between water waves and floating freshwater
  ice.
\newblock \emph{Physics of Fluids}, 30\penalty0 (9), 2018.

\bibitem[Passerotti et~al.(2022)Passerotti, Bennetts, von Bock~und Polach,
  Alberello, Puolakka, Dolatshah, Monbaliu, and
  Toffoli]{passerotti2022interactions}
G.~Passerotti, L.~G. Bennetts, F.~von Bock~und Polach, A.~Alberello,
  O.~Puolakka, A.~Dolatshah, J.~Monbaliu, and A.~Toffoli.
\newblock Interactions between irregular wave fields and sea ice: {A} physical
  model for wave attenuation and ice breakup in an ice tank.
\newblock \emph{Journal of Physical Oceanography}, 52\penalty0 (7):\penalty0
  1431--1446, 2022.

\bibitem[Meylan et~al.(2018)Meylan, Bennetts, Mosig, Rogers, Doble, and
  Peter]{meylan2018dispersion}
M.~H. Meylan, L.~G. Bennetts, J.~Mosig, W.~Rogers, M.~Doble, and M.~A. Peter.
\newblock Dispersion relations, power laws, and energy loss for waves in the
  marginal ice zone.
\newblock \emph{Journal of Geophysical Research: Oceans}, 123\penalty0
  (5):\penalty0 3322--3335, 2018.

\bibitem[Weber(1987)]{weber1987wave}
J.~E. Weber.
\newblock Wave attenuation and wave drift in the marginal ice zone.
\newblock \emph{Journal of Physical Oceanography}, 17\penalty0 (12):\penalty0
  2351--2361, 1987.

\bibitem[Keller(1998)]{keller1998gravity}
J.~B. Keller.
\newblock Gravity waves on ice-covered water.
\newblock \emph{Journal of Geophysical Research: Oceans}, 103\penalty0
  (C4):\penalty0 7663--7669, 1998.

\bibitem[Liu and Mollo-Christensen(1988)]{liu1988wave}
A.~K. Liu and E.~Mollo-Christensen.
\newblock Wave propagation in a solid ice pack.
\newblock \emph{Journal of Physical Oceanography}, 18\penalty0 (11):\penalty0
  1702--1712, 1988.

\bibitem[Wang and Shen(2010)]{wang2010gravity}
R.~Wang and H.~H. Shen.
\newblock Gravity waves propagating into an ice-covered ocean: {A} viscoelastic
  model.
\newblock \emph{Journal of Geophysical Research: Oceans}, 115\penalty0 (C6),
  2010.

\bibitem[Zhao et~al.(2017)Zhao, Cheng, and Shen]{zhao2017nature}
X.~Zhao, S.~Cheng, and H.~H. Shen.
\newblock Nature of wave modes in a coupled viscoelastic layer over water.
\newblock \emph{Journal of Engineering Mechanics}, 143\penalty0 (10):\penalty0
  04017114, 2017.

\bibitem[Mosig et~al.(2015)Mosig, Montiel, and Squire]{mosig2015comparison}
J.~E. Mosig, F.~Montiel, and V.~A. Squire.
\newblock Comparison of viscoelastic-type models for ocean wave attenuation in
  ice-covered seas.
\newblock \emph{Journal of Geophysical Research: Oceans}, 120\penalty0
  (9):\penalty0 6072--6090, 2015.

\bibitem[De~Santi and Olla(2017)]{de2017effect}
F.~De~Santi and P.~Olla.
\newblock Effect of small floating disks on the propagation of gravity waves.
\newblock \emph{Fluid Dynamics Research}, 49\penalty0 (2):\penalty0 025512,
  2017.

\bibitem[Alberello et~al.(2019)Alberello, Onorato, Bennetts, Vichi, Eayrs,
  MacHutchon, and Toffoli]{alberello2019brief}
A.~Alberello, M.~Onorato, L.~Bennetts, M.~Vichi, C.~Eayrs, K.~MacHutchon, and
  A.~Toffoli.
\newblock Pancake ice floe size distribution during the winter expansion of the
  {A}ntarctic marginal ice zone.
\newblock \emph{The Cryosphere}, 13\penalty0 (1):\penalty0 41--48, 2019.

\bibitem[De~Santi et~al.(2018)De~Santi, De~Carolis, Olla, Doble, Cheng, Shen,
  Wadhams, and Thomson]{de2018ocean}
F.~De~Santi, G.~De~Carolis, P.~Olla, M.~Doble, S.~Cheng, H.~H. Shen,
  P.~Wadhams, and J.~Thomson.
\newblock On the ocean wave attenuation rate in grease-pancake ice, a
  comparison of viscous layer propagation models with field data.
\newblock \emph{Journal of Geophysical Research: Oceans}, 123\penalty0
  (8):\penalty0 5933--5948, 2018.

\bibitem[Wadhams(1973)]{wadhams1973attenuation}
P.~Wadhams.
\newblock Attenuation of swell by sea ice.
\newblock \emph{Journal of Geophysical Research}, 78\penalty0 (18):\penalty0
  3552--3563, 1973.

\bibitem[Shen and Squire(1998)]{shen1998wave}
H.~H. Shen and V.~A. Squire.
\newblock Wave damping in compact pancake ice fields due to interactions
  between pancakes.
\newblock \emph{Antarctic Sea Ice: Physical Processes, Interactions and
  Variability}, 74:\penalty0 325--341, 1998.

\bibitem[Kohout et~al.(2011)Kohout, Meylan, and Plew]{kohout2011wave}
A.~L. Kohout, M.~H. Meylan, and D.~R. Plew.
\newblock Wave attenuation in a marginal ice zone due to the bottom roughness
  of ice floes.
\newblock \emph{Annals of Glaciology}, 52\penalty0 (57):\penalty0 118--122,
  2011.

\bibitem[Squire(2018)]{squire2018fresh}
V.~A. Squire.
\newblock A fresh look at how ocean waves and sea ice interact.
\newblock \emph{Philosophical Transactions of the Royal Society A},
  376\penalty0 (2129):\penalty0 20170342, 2018.

\bibitem[Boutin et~al.(2018)Boutin, Ardhuin, Dumont, S{\'e}vigny,
  Girard-Ardhuin, and Accensi]{boutin2018floe}
G.~Boutin, F.~Ardhuin, D.~Dumont, C.~S{\'e}vigny, F.~Girard-Ardhuin, and
  M.~Accensi.
\newblock Floe size effect on wave-ice interactions: Possible effects,
  implementation in wave model, and evaluation.
\newblock \emph{Journal of Geophysical Research: Oceans}, 123\penalty0
  (7):\penalty0 4779--4805, 2018.

\bibitem[Fox and Squire(1991)]{fox1991coupling}
C.~Fox and V.~A. Squire.
\newblock Coupling between the ocean and an ice shelf.
\newblock \emph{Annals of Glaciology}, 15:\penalty0 101--108, 1991.

\bibitem[Balmforth and Craster(1999)]{balmforth1999ocean}
N.~Balmforth and R.~Craster.
\newblock Ocean waves and ice sheets.
\newblock \emph{Journal of Fluid Mechanics}, 395:\penalty0 89--124, 1999.

\bibitem[Kalyanaraman et~al.(2019)Kalyanaraman, Bennetts, Lamichhane, and
  Meylan]{kalyanaraman2019shallow}
B.~Kalyanaraman, L.~G. Bennetts, B.~Lamichhane, and M.~H. Meylan.
\newblock On the shallow-water limit for modelling ocean-wave induced ice-shelf
  vibrations.
\newblock \emph{Wave Motion}, 90:\penalty0 1--16, 2019.

\bibitem[Holdsworth and Glynn(1981)]{holdsworth1981mechanism}
G.~Holdsworth and J.~Glynn.
\newblock A mechanism for the formation of large icebergs.
\newblock \emph{Journal of Geophysical Research: Oceans}, 86\penalty0
  (C4):\penalty0 3210--3222, 1981.

\bibitem[Holdsworth and Glynn(1978)]{holdsworth1978iceberg}
G.~Holdsworth and J.~Glynn.
\newblock Iceberg calving from floating glaciers by a vibrating mechanism.
\newblock \emph{Nature}, 274\penalty0 (5670):\penalty0 464--466, 1978.

\bibitem[Sergienko(2013)]{sergienko2013normal}
O.~V. Sergienko.
\newblock Normal modes of a coupled ice-shelf/sub-ice-shelf cavity system.
\newblock \emph{Journal of Glaciology}, 59\penalty0 (213):\penalty0 76--80,
  2013.

\bibitem[Meylan et~al.(2017)Meylan, Bennetts, Hosking, and
  Catt]{meylan2017calculation}
M.~H. Meylan, L.~G. Bennetts, R.~J. Hosking, and E.~Catt.
\newblock On the calculation of normal modes of a coupled
  ice-shelf/sub-ice-shelf cavity system.
\newblock \emph{Journal of Glaciology}, 63\penalty0 (240):\penalty0 751--754,
  2017.

\bibitem[Papathanasiou et~al.(2019)Papathanasiou, Karperaki, and
  Belibassakis]{papathanasiou2019resonant}
T.~K. Papathanasiou, A.~E. Karperaki, and K.~A. Belibassakis.
\newblock On the resonant hydroelastic behaviour of ice shelves.
\newblock \emph{Ocean Modelling}, 133:\penalty0 11--26, 2019.

\bibitem[Bennetts and Meylan(2021)]{bennetts2021complex}
L.~G. Bennetts and M.~H. Meylan.
\newblock Complex resonant ice shelf vibrations.
\newblock \emph{SIAM Journal on Applied Mathematics}, 81\penalty0 (4):\penalty0
  1483--1502, 2021.

\bibitem[Vinogradov and Holdsworth(1985)]{vinogradov1985oscillation}
O.~Vinogradov and G.~Holdsworth.
\newblock Oscillation of a floating glacier tongue.
\newblock \emph{Cold Regions Science and Technology}, 10\penalty0 (3):\penalty0
  263--271, 1985.

\bibitem[Ilyas et~al.(2018)Ilyas, Meylan, Lamichhane, and
  Bennetts]{ilyas2018time}
M.~Ilyas, M.~H. Meylan, B.~Lamichhane, and L.~G. Bennetts.
\newblock Time-domain and modal response of ice shelves to wave forcing using
  the finite element method.
\newblock \emph{Journal of Fluids and Structures}, 80:\penalty0 113--131, 2018.

\bibitem[Meylan et~al.(2021)Meylan, Ilyas, Lamichhane, and
  Bennetts]{meylan2021swell}
M.~H. Meylan, M.~Ilyas, B.~P. Lamichhane, and L.~G. Bennetts.
\newblock Swell-induced flexural vibrations of a thickening ice shelf over a
  shoaling seabed.
\newblock \emph{Proceedings of the Royal Society A}, 477\penalty0
  (2254):\penalty0 20210173, 2021.

\bibitem[Liang et~al.(2024)Liang, Pitt, and Bennetts]{liang2024pan}
J.~Liang, J.~P. Pitt, and L.~G. Bennetts.
\newblock Pan-{A}ntarctic assessment of ice shelf flexural responses to ocean
  waves.
\newblock \emph{Journal of Geophysical Research: Oceans}, 129\penalty0
  (8):\penalty0 e2023JC020824, 2024.

\bibitem[Freed-Brown et~al.(2012)Freed-Brown, Amundson, MacAyeal, and
  Zhang]{freed2012blocking}
J.~Freed-Brown, J.~M. Amundson, D.~R. MacAyeal, and W.~W. Zhang.
\newblock Blocking a wave: frequency band gaps in ice shelves with periodic
  crevasses.
\newblock \emph{Annals of glaciology}, 53\penalty0 (60):\penalty0 85--89, 2012.

\bibitem[Nekrasov and MacAyeal(2023)]{nekrasov2023ocean}
P.~Nekrasov and D.~R. MacAyeal.
\newblock Ocean wave blocking by periodic surface rolls fortifies {A}rctic ice
  shelves.
\newblock \emph{Journal of Glaciology}, 69\penalty0 (278):\penalty0 1740--1750,
  2023.

\bibitem[Kalyanaraman et~al.(2021)Kalyanaraman, Meylan, Lamichhane, and
  Bennetts]{kalyanaraman2021icefem}
B.~Kalyanaraman, M.~H. Meylan, B.~P. Lamichhane, and L.~G. Bennetts.
\newblock {iceFEM: A} {FreeFem} package for wave induced ice-shelf vibrations.
\newblock \emph{Journal of Open Source Software}, 6\penalty0 (59):\penalty0
  2939, 2021.

\bibitem[Bennetts et~al.(2022{\natexlab{c}})Bennetts, Liang, and
  Pitt]{bennetts2022modeling}
L.~G. Bennetts, J.~Liang, and J.~P. Pitt.
\newblock Modeling ocean wave transfer to {Ross Ice Shelf} flexure.
\newblock \emph{Geophysical Research Letters}, 49\penalty0 (21):\penalty0
  e2022GL100868, 2022{\natexlab{c}}.

\bibitem[Papathanasiou and Belibassakis(2019)]{papathanasiou2019nonconforming}
T.~K. Papathanasiou and K.~A. Belibassakis.
\newblock A nonconforming hydroelastic triangle for ice shelf modal analysis.
\newblock \emph{Journal of Fluids and Structures}, 91:\penalty0 102741, 2019.

\bibitem[Sergienko(2017)]{sergienko2017behavior}
O.~V. Sergienko.
\newblock Behavior of flexural gravity waves on ice shelves: Application to the
  {Ross Ice Shelf}.
\newblock \emph{Journal of Geophysical Research: Oceans}, 122\penalty0
  (8):\penalty0 6147--6164, 2017.

\bibitem[Tazhimbetov et~al.(2023)Tazhimbetov, Almquist, Werpers, and
  Dunham]{tazhimbetov2023simulation}
N.~Tazhimbetov, M.~Almquist, J.~Werpers, and E.~M. Dunham.
\newblock Simulation of flexural-gravity wave propagation for elastic plates in
  shallow water using an energy-stable finite difference method with weakly
  enforced boundary and interface conditions.
\newblock \emph{Journal of Computational Physics}, 493:\penalty0 112470, 2023.

\bibitem[Sergienko(2010)]{sergienko2010elastic}
O.~V. Sergienko.
\newblock Elastic response of floating glacier ice to impact of long-period
  ocean waves.
\newblock \emph{Journal of Geophysical Research: Earth Surface}, 115\penalty0
  (F4), 2010.

\bibitem[Kalyanaraman et~al.(2020)Kalyanaraman, Meylan, Bennetts, and
  Lamichhane]{kalyanaraman2020coupled}
B.~Kalyanaraman, M.~H. Meylan, L.~G. Bennetts, and B.~P. Lamichhane.
\newblock A coupled fluid-elasticity model for the wave forcing of an
  ice-shelf.
\newblock \emph{Journal of Fluids and Structures}, 97:\penalty0 103074, 2020.

\bibitem[Abrahams et~al.(2023)Abrahams, Mierzejewski, Dunham, and
  Bromirski]{abrahams2023ocean}
L.~Abrahams, J.~Mierzejewski, E.~Dunham, and P.~D. Bromirski.
\newblock Ocean surface gravity wave excitation of flexural gravity and
  extensional {L}amb waves in ice shelves.
\newblock \emph{Seismica}, 2\penalty0 (1), 2023.

\bibitem[Bennetts et~al.(2024{\natexlab{b}})Bennetts, Williams, and
  Porter]{bennetts2024thin}
L.~G. Bennetts, T.~D. Williams, and R.~Porter.
\newblock A thin-plate approximation for ocean wave interactions with an ice
  shelf.
\newblock \emph{Journal of Fluid Mechanics}, 984:\penalty0 A48,
  2024{\natexlab{b}}.

\bibitem[Alberello et~al.(2022)Alberello, Bennetts, Onorato, Vichi, MacHutchon,
  Eayrs, Ntamba, Benetazzo, Bergamasco, Nelli, et~al.]{alberello2022three}
A.~Alberello, L.~G. Bennetts, M.~Onorato, M.~Vichi, K.~MacHutchon, C.~Eayrs,
  B.~N. Ntamba, A.~Benetazzo, F.~Bergamasco, F.~Nelli, et~al.
\newblock Three-dimensional imaging of waves and floes in the marginal ice zone
  during a cyclone.
\newblock \emph{Nature communications}, 13\penalty0 (1):\penalty0 4590, 2022.

\bibitem[Greene et~al.(2022)Greene, Gardner, Schlegel, and
  Fraser]{greene2022antarctic}
C.~A. Greene, A.~S. Gardner, N.-J. Schlegel, and A.~D. Fraser.
\newblock Antarctic calving loss rivals ice-shelf thinning.
\newblock \emph{Nature}, 609\penalty0 (7929):\penalty0 948--953, 2022.

\bibitem[Paolo et~al.(2015)Paolo, Fricker, and Padman]{paolo2015volume}
F.~S. Paolo, H.~A. Fricker, and L.~Padman.
\newblock Volume loss from {A}ntarctic ice shelves is accelerating.
\newblock \emph{Science}, 348\penalty0 (6232):\penalty0 327--331, 2015.

\bibitem[Bassis et~al.(2024)Bassis, Crawford, Kachuck, Benn, Walker, Millstein,
  Duddu, {\AA}str{\"o}m, Fricker, and Luckman]{bassis2024stability}
J.~N. Bassis, A.~Crawford, S.~B. Kachuck, D.~I. Benn, C.~Walker, J.~Millstein,
  R.~Duddu, J.~{\AA}str{\"o}m, H.~Fricker, and A.~Luckman.
\newblock Stability of ice shelves and ice cliffs in a changing climate.
\newblock \emph{Annual Review of Earth and Planetary Sciences}, 52, 2024.

\bibitem[Dumont(2022)]{dumont2022marginal}
D.~Dumont.
\newblock Marginal ice zone dynamics: history, definitions and research
  perspectives.
\newblock \emph{Philosophical Transactions of the Royal Society A},
  380\penalty0 (2235):\penalty0 20210253, 2022.

\bibitem[Teder et~al.(2022)Teder, Bennetts, Reid, and Massom]{teder2022sea}
N.~J. Teder, L.~G. Bennetts, P.~A. Reid, and R.~A. Massom.
\newblock Sea ice-free corridors for large swell to reach {A}ntarctic ice
  shelves.
\newblock \emph{Environmental Research Letters}, 17\penalty0 (4):\penalty0
  045026, 2022.

\bibitem[Teder et~al.(2025)Teder, Bennetts, Reid, Massom, Pitt, Scambos, and
  Fraser]{teder2025large}
N.~J. Teder, L.~G. Bennetts, P.~A. Reid, R.~A. Massom, J.~P. Pitt, T.~A.
  Scambos, and A.~D. Fraser.
\newblock Large-scale ice shelf calving events follow prolonged amplifications
  in flexure.
\newblock \emph{Nature Geosciences}, page in press, 2025.

\bibitem[Greene et~al.(2018)Greene, Young, Gwyther, Galton-Fenzi, and
  Blankenship]{greene2018seasonal}
C.~A. Greene, D.~A. Young, D.~E. Gwyther, B.~K. Galton-Fenzi, and D.~D.
  Blankenship.
\newblock Seasonal dynamics of {Totten Ice Shelf} controlled by sea ice
  buttressing.
\newblock \emph{The Cryosphere}, 12\penalty0 (9):\penalty0 2869--2882, 2018.

\end{thebibliography}

\end{document}